# Entanglement and weak interaction driven mobility of small molecules in polymer networks

Rajarshi Guha[a,*] Subhadip Ghosh[c], Darrell Velegol[a], Peter J. Butler[b], Ayusman Sen[c], Jennifer L. Ross[d]

[a]Department of Chemical Engineering, Pennsylvania State University,
University Park, Pennsylvania, 16802, USA

[b]Department of Biomedical Engineering, Pennsylvania State University,
University Park, Pennsylvania, 16802, USA

[c]Department of Chemistry, Pennsylvania State University,
University Park, Pennsylvania, 16802, USA

[d]Department of Physics, Syracuse University,
Syracuse, NY 13244

[*]Correspondence should be addressed to R.G. (rajarshiche@gmail.com)

## ABSTRACT

Diffusive transport of small molecules within the internal structures of biological and synthetic material systems is complex because the crowded environment presents chemical and physical barriers to mobility. We explored this mobility using a synthetic experimental system of small dye molecules diffusing within a polymer network at short time scales. We find that the diffusion of inert molecules is inhibited by the presence of the polymers. Counter-intuitively, small, hydrophobic molecules display smaller reduction in mobility and also able to diffuse faster through the system by leveraging crowding specific parameters. We explained this phenomenon by developing a *de novo* model and using these results, we hypothesized that non-specific hydrophobic interactions between the molecules and polymer chains could localize the molecules into compartments of overlapped and entangled chains where they experience microviscosity, rather than macroviscosity. We introduced a characteristic interaction time parameter to quantitatively explain experimental results in the light of frictional effects and molecular interactions. Our model is in good agreement with the experimental results and allowed us to classify molecules into two different mobility categories solely based on interaction. By changing the surface group, polymer molecular weight, and by adding salt to the medium, we could further modulate the mobility and mean square displacements of interacting molecules. Our work has implications in understanding intracellular diffusive transport in microtubule networks and other systems with macromolecular crowding and could lead to transport enhancement in synthetic polymer systems.

**TOC graphic**

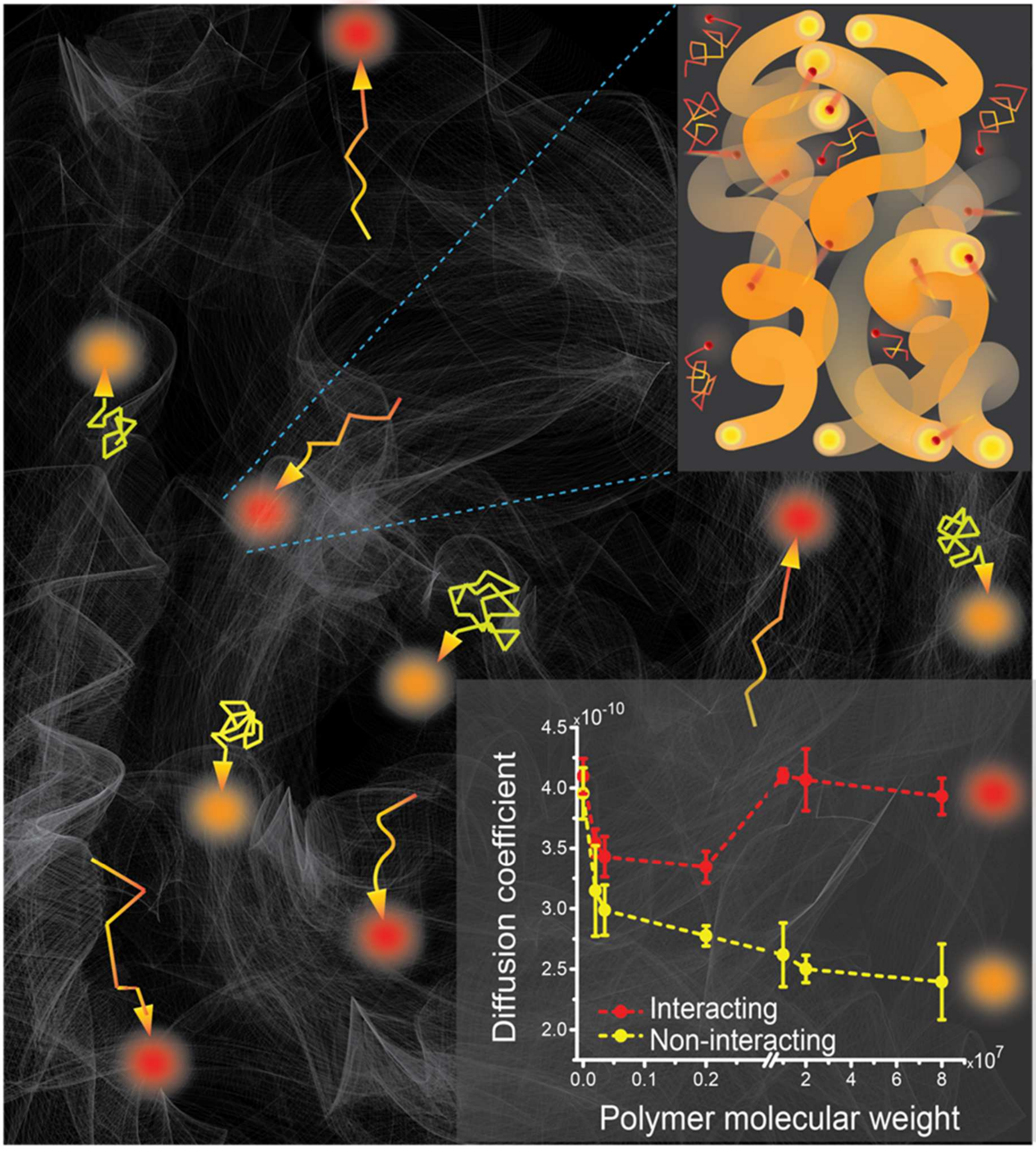

x10-10
4.5
4.0
3.5
3.0
2.5
2.0
Diffusion coefficient
Interacting
Non-interacting
0.0
0.1
0.2
2
4
6
8
x107
Polymer molecular weight

## Introduction:

Diffusion is a fundamental process to transport species, signals, and information in materials and biological systems. Understanding diffusion in complex polymeric systems has significant practical implications for polymer characterization, drug delivery, cell biology, molecular detection, rechargeable batteries etc. In intracellular, biological system, the diffusion of small molecules dictates the dynamics, the temporal and spatial scales, of cellular processes as fast as millisecond timescales[1]. The chemical and physical properties of the polymer networks may alter the diffusion of small molecules by either hindering or enhancing the diffusivity through facilitated diffusion[2–5] or active transport[6–8]. Mobility of small molecules is thought to follow Stokes-Einstein relation in polymer medium and is believed to get minimally affected despite the presence of polymers[9]. Additionally, the underlying mechanisms of diffusion of small molecules in macromolecular crowding, as opposed to diffusion of larger sized colloidal particles comparable to the length scale of the system [9–12], is not well understood.

Transport in polymer systems is known to be dependent on length scales, but for small molecules, it is also dependent on local viscosity or microviscosity[13,14] and interactions in crowded medium [15,16]. The non-specific or weak interaction is a typical hallmark of small molecules in crowded biological milieu [16] which itself dissuades stronger, specific interactions[17]. These weak interactions enable *in vitro* chemotaxis of small molecules in polymer gradients[18] and could be responsible for anomalous diffusion[15]. The effects of interactions could manifest in nonmonotonic mobility patterns with a change in the degree of crowding [19,20]. However, there is an apparent gap in understanding of crowding and interactions in the context of diffusion of small molecules, taking into account the spatio-temporal scales, molecular friction, and microviscosity of the local environment[21,22]. Overall, we seek to understand how the presence and absence of weak interactions subject to the microviscous environment change the diffusion landscape of small molecules in macromolecular crowding.

Structure-function of intracellular cytoskeletal networks is seemingly important in the context of reactive-diffusive organization of these components, which is dictated by diffusive interactions [23,24]. Particularly important are the cases of microtubule networks (MTs) [25] and membraneless liquid-liquid droplet separation (LLPS) systems [26], where the spatial organization of constituents forms certain mesh sizes with more liquid-like healthy states and more solid-like diseased states[27,28]. It is thus expected that the internal network structures of these condensates organize accordingly to perform required functional activities, including molecular mobility in the systems.

To directly explore the effects of polymer networks on small molecule diffusion, we performed a series of experiments examining the ability of sticky or inert small molecules diffusing in submillisecond time scales within physical networks of synthetic polymers in homogeneous aqueous solutions. We used polymer networks composed of physical crosslinks of neutral polymer chains. We found that the mobility of inert molecules is hindered by the polymer network with increasing viscosity of medium and polymer molecular weights, as expected. Conversely, sticky hydrophobic molecules and small nanoparticles, due to weak interactions with the polymer network, displayed altered diffusion in solutions and a faster general diffusion coefficient through internalization within polymer network structures. These sticky molecules, analogous to intelligent systems, can sense and explore their local environment. Surprisingly, at short time scales and within entangled polymer compartments, such interacting or sticky molecules barely experience the presence of polymers and their mean square displacements suggest that the molecules diffuse in dilute solutions.  These experimental results were simulated with models described by different local viscosities within the network, which were due to the inherent structural length scale of the polymer solution and non-specific interactions between the polymer network and the small molecule. Importantly, we find that the mobility characteristics of sticky molecules could be further enhanced at higher salt concentration and optimized polymer molecular weight due to stronger interaction parameters. Contrary to mobility dampening due to interactions, we found that polymer networks, through appropriate crowding conditions, actually could enhance mobility depending on the nature of the molecule. Projecting these observations, theoretically, on biopolymer networks, we found structural evidence that MT networks within healthy cells maintain very specific crowding conditions as opposed to diseased cells, and this condition is appeared to be evolutionary conserved. We describe an experimental model system to differentiate the extent of non-specific molecular interaction between polymer network and small molecules and thereby, quantify molecular characteristics of the system through diffusivity measurements.

## Results:

### Transport patterns of similar sized small molecules are fundamentally different in polymer networks.

First, we needed to screen out molecular systems of similar size and desired interaction properties which we can compare for further study and we used a 3-channel microfluidic setup for this purpose (SI, Figure S1). Specifically, we have previously quantified the “chemotaxis index” (CI) of different small molecules diffusing in gradients of polymer networks in microfluidic channels [18]. Molecular chemotaxis defines the ability of certain molecules to crawl along the increasing concentrations of polymer and CI denotes an empirical index correlating molecular movements in concentration gradients to molecular functional groups and charges. From those prior results, we know that some molecules can be weakly or non-interacting with the polymer network, such as 6-HEX, which has a CI = 0, and other molecules are more interacting, such as Rhodamine6G (Rh6G), which has a CI = 6. Supporting this observation, in gradients of polymer and under similar conditions in microfluidic channels, Rh6G molecules showed chemotaxis several-fold larger than 6-HEX molecules (SI, Figure S1). The experimental results are briefly described in the SI. We found 4-fold larger displacements of Rh6G molecule than 6-HEX in microfluidic channels under the same conditions. In structural terms, the hydrophobic alkyl groups might be responsible for such interaction, which are present in single molecules such as Rh6G. Since concentration gradients can cause enhanced molecular transport in the presence of interaction, we hypothesize that interaction can also cause enhanced diffusive mobility in homogeneous solution as well, in comparison to control cases. Overall, the microfluidic experiments helped to screen out the basic molecular systems to study with a structure-driven hypothesis.

We have used polyethylene oxide (PEO) solution, a well-characterized polymer network in an aqueous environment. We have studied different molecular weights (MWs) of PEO, but at the same wt% to quantify the variation of diffusion coefficients with spatial scales of the polymer network by keeping the number of potential interaction sites the same across all MWs. To measure the diffusion, we mostly used fluorescence correlation spectroscopy (FCS), which enables determination of the diffusion coefficient from the intensity fluctuations of fluorophores (Figure 1A) as they pass through a diffraction-limited confocal volume within the sample.

We find that in dilute polymer solutions, both inert (non-interacting) and sticky (interacting) molecules behave in similar ways and their diffusive mobility decreases with an increase in polymer molecular weight as expected from the increase in local friction factor dictated by microviscosity of the local environment. However, in an entangled polymer solution in the semidilute regime, the presence or absence of molecular interactions between the diffusing molecule and the network could cause enhanced or reduced diffusion (Figure 1B).

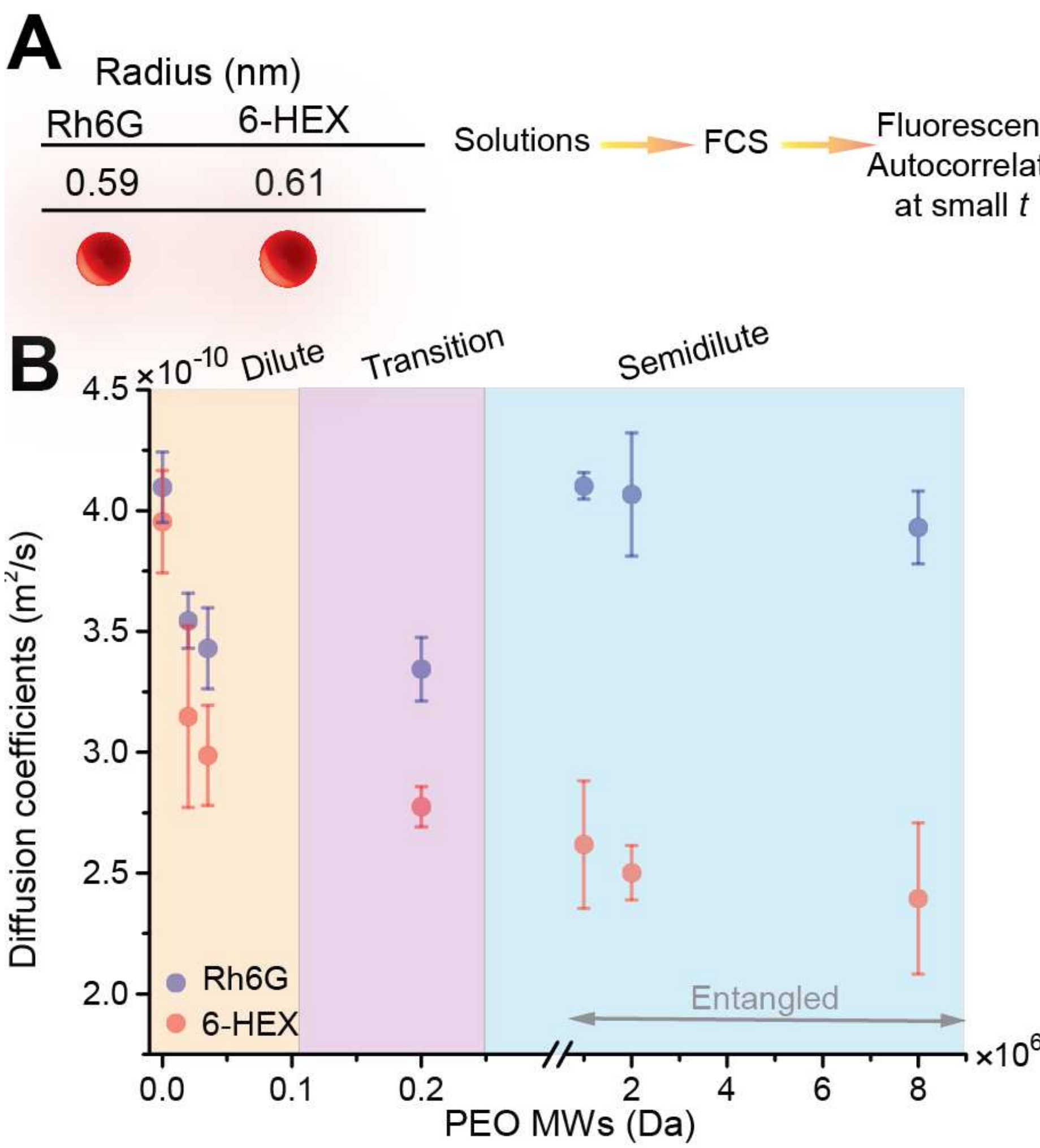

**Figure 1. The diffusion of small molecules in semidilute entangled polymer solutions depends on molecular interactions.** The diffusivity pattern of an inert molecule (6-HEX) monotonically decreases as

opposed to the diffusivity pattern of a sticky molecule (Rh6G) under the same conditions. (A) The equivalence of molecular size between Rh6G and 6-HEX is not manifested in their diffusion behavior as measured using FCS analysis at small time intervals. (B) Measured diffusion coefficients of 6-HEX (blue) at submillisecond time scale in 0.5 wt% PEO solutions at increasing MWs show a decreasing pattern depending on the number of monomers and concentration of the polymer present. Measured diffusion coefficients of Rh6G (red) at submillisecond time scale initially decrease in the dilute regime of polymer concentration, but show a relative increase in the semi-dilute entangled regime of respective MWs (≥ 200 kDa ) of the polymer. Notice that the diffusion coefficient of Rh6G in the solution of $10^6$ PEO MW is similar to DI water value. The error bars are standard deviation from at least 5 independent measurements.

**Small molecules do not follow Stokes-Einstein diffusion in polymer networks.**

The classical Stokes-Einstein formulation of Brownian diffusion, owing to its assumption of non-interacting particles, depends only on the balance between thermal fluctuations and solvent friction based on the bulk or macroviscosity of the environment [29,30]. The Stokes-Einstein diffusion coefficient ($D_{SE}$) of a molecule of radius $a_m$ at temperature $T$ in a liquid of bulk viscosity $\eta$ is given by $D_{SE} = K_B T/\gamma$, where the friction factor $\gamma = 6\pi\eta a_m$, and $K_B$ is the Boltzmann constant. These assumptions ignore other thermal and non-thermal forces as well as characteristic length scales prevalent in real systems [6–8,31–33]. Additionally, in cases when small molecules are well enclosed within polymer tube structures *i.e.* $2a_m \ll a_t$, where $a_t$ is the virtual tube diameter due to chain entanglement, deviation of a molecule's behavior from Stokes-Einstein theory is quite prominent owing to its sensitivity on microviscosity [34], instead of macroviscosity or bulk viscosity of the solution. Here, microviscosity refers to the viscosity sensed by solute molecules, like Rh6G or 6-HEX, with sizes much smaller than the radius of gyration of the polymer. Therefore, the ratio of experimentally measured diffusion coefficient ($D_{expt}$) to $D_{SE}$ based on solution macroviscosity exponentially increases due to increase in PEO MWs at the same concentration (SI, Figure S2).

In the case of colloidal particles following Stokes-Einstein diffusion coefficient and mean square displacements proportional to the coefficient, the normalized diffusion coefficient ($D_{expt}/D_0$) should linearly increase with an inverse of solution macroviscosity ($1/\eta$) and the distribution of the spatio-temporal positions of particles should be Gaussian. We performed experiments with 100 nm APSL particles diffusing in 0.3% PEO solutions of different MWs and found that the viscosity scaling follows Stokes-Einstein behavior and the autocorrelation function which dictates the spatio-temporal distribution, was indeed Gaussian (SI, Figures S3A and S3B).

The diffusion coefficient of inert molecules (6-HEX), as expected, is affected by the viscosity of the polymer solution, and therefore, decreases with increasing MWs of polymers in solution at the same concentration of 0.5 wt% (Figure 1B). However, we find that Stokes-Einstein theory cannot explain the diffusion behavior of inert molecules (SI, Figure S2). Conversely, we find that the diffusion coefficient is non-monotonic for sticky molecules (Rh6G), even if it is of similar size to the inert molecule (Figure 1A) and the enhancement occurs despite the increase in viscosity of polymer solution with the increase in MWs. Interestingly, in the early entangled regime and at small time intervals, a sticky molecule does not feel the presence of the polymer and diffuses as if the medium is DI water as evident from the diffusion coefficient of Rh6G in PEO solution of MW $10^6$. Both the patterns of normalized diffusivity for sticky and inert molecules show similar trends at lower MWs of PEO solution (dilute regime), but diverge at higher MWs (semidilute regime). This separation of mobility pattern starts with the initiation of transition to the semidilute regime of the polymer solution (~ MW of $2 \times 10^5$). This observation further implies that a fundamentally different mechanism is at play in the case of sticky molecules in discussed polymer networks.

**Concentration and correlation length of the polymer network dictate the mobility of inert, non-interacting small molecules.**

We modeled the mobility of inert molecules diffusing in a solution with a certain number of monomer blobs according to one-dimensional obstruction theory as detailed in the supplementary information (SI, section S2 and Tables S1-S2). As per the theory, an inert molecule exhibits a monotonous decrease in diffusion coefficients ($D_x$) with the increase in PEO MWs following: $D_x = D_0 \, e^{-N/N_c}$, where $D_0$ is the diffusion coefficient of the molecule in DI water, $N$ is the number of PEO monomers in one MW of the polymer and $N_c$ is the characteristic number of monomer units which causes an exponential decrease in molecular diffusivity. Measurement of diffusion coefficients of inert 6-HEX molecules in PEO solutions shows that the theory supports experiments (Figure 2A). We extended our experimental system to another inert molecule, Rhodamine B tagged polyethylene glycol of MW of $10^3$ (denoted as "RhB – PEG"), which is larger than 6-HEX, but still satisfies the condition: $2a_m \ll a_t$. We hypothesize that RhB –PEG would not interact with PEO segments because it consists of similar structural units as PEO chains. Since this macromolecule is similar to a small particle diffusing in polymer solution randomly, it would be obstructed by the increasing number of monomers in the crowded milieu. After performing diffusion coefficient measurements under the same conditions, we found that RhB –PEG indeed follows the mobility pattern of 6-HEX molecules (Figure 2B).

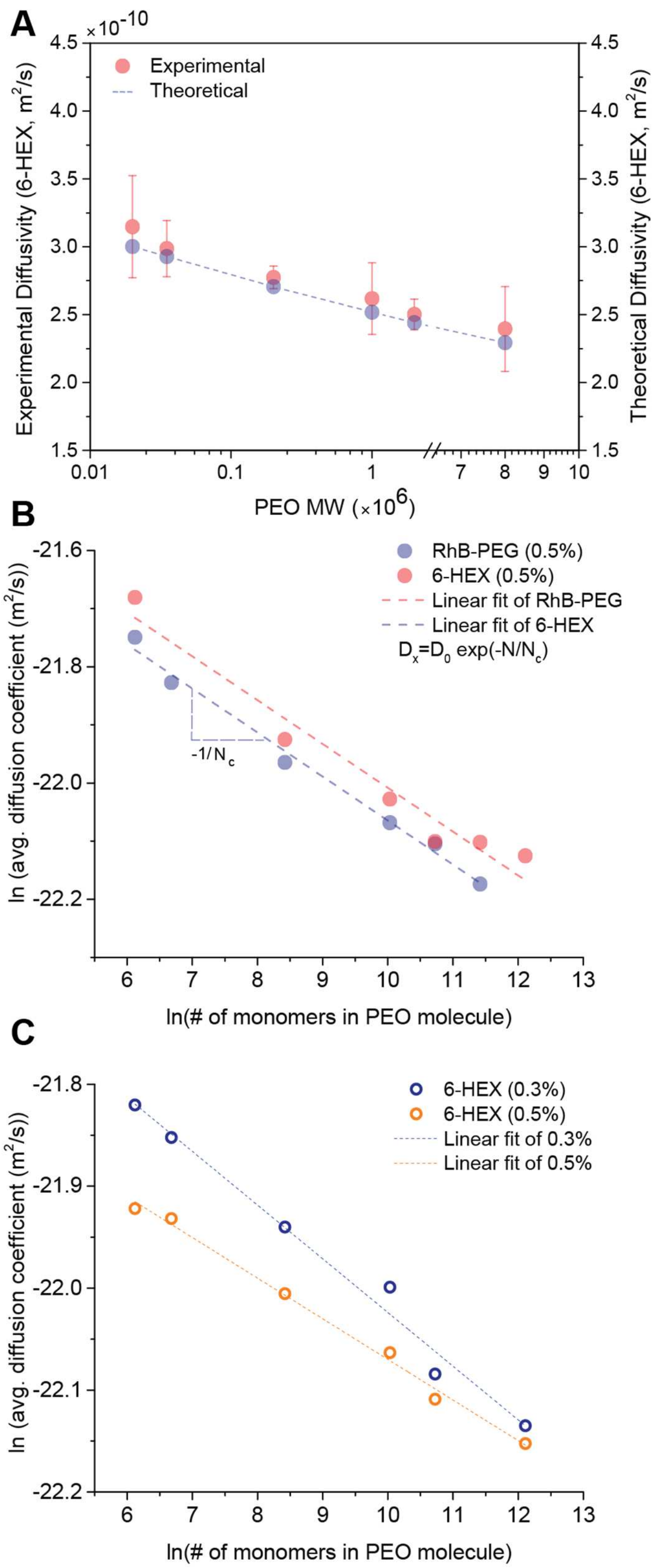

**Figure 2. The diffusion coefficient of inert molecules decreases with an increase in MW of the polymer**. (A) Experimental measurements and modeled diffusion coefficients of inert 6-HEX molecule in 0.5 wt% PEO solutions with x-axis in log scale. (B) Natural logarithm of average diffusivity of inert 6-HEX and RhB - PEG molecules in 0.5 wt% PEO solution shows a linear pattern with similar slopes ($-1/N_c$) with an increase in the number of monomers of the polymer for different MW expressed in the log scale. (C) The natural logarithm of average diffusivity of inert 6-HEX molecules in 0.3 wt% and 0.5 wt% PEO solutions at different MWs shows different slopes with an increase in the number of monomers in MW scale of the polymer expressed in log scale. We plotted the diffusivity correlation $D_x = D_0\ e^{-N/N_c}$ with x-axis: $\ln(N)$ and y-axis: $\ln(D_x)$. Error bars representing standard deviations and average values used in the figures are calculated using at least 5 independent measurements. For error bars on log-log plots and theoretical details, please see SI section S2, Figures S4 and S5.

From the commonality of slopes and therefore, the same $N_c$ values between 6-HEX and RhB-PEG (Figure 2B and SI, Figures S4 and S5), we conclude that $N_c$ suggests no general dependency on the type of diffusing, inert small molecule.

Following the one-dimensional obstruction model of inert molecules, the theoretical number of compartments available to the molecule would be $N/N_c$ at a certain concentration. This number increases with molecular weights and leads to the creation of even smaller size compartments with the result of increased resistance in the medium. This is because previously available free space gets further divided at higher MWs resulting in inhibition of mobility of inert molecules. In other words, the effective friction factor ($\zeta$) of inert molecules would scale as $\zeta \sim e^{N/N_c}$ and would lead to an exponential decrease in mobility. Therefore, in parallel to the obstruction theory, we can describe the mobility pattern of inert molecules with an effective compartment model.

It is expected that $N_c$ defines some type of physical length scale of the polymer network which the diffusing molecule experiences as a hindrance. By performing diffusion coefficient measurements of 6-HEX at different concentrations (eg. 0.3% and 0.5% PEO MWs), we found that the rate of decrease of diffusion coefficient with the increase in MWs (or slope: $-1/N_c$) is smaller at higher concentrations indicating a scaling $N_c \sim C^{\nu}$ (Figure 2C and SI, Table S2). This correlation supports the hydrodynamic scaling model by Phillies where the exponent $\nu$ can be between 0.5 – 1[35]. In our system concerning small molecules, we obtained the best fit to experimental data when $\nu$ is the Flory exponent in good solvent (SI, Table S2). RhB-PEG also follows scaling similar to 6-HEX. This further suggests that at the same MW, but at different concentrations, $N_c \sim \xi^{1-3\nu}$ , where correlation length is $\xi$ (SI, Table S2). Therefore, such scaling analysis might provide an easier way to quantify Flory exponent or correlation lengths of the system by simply measuring diffusion coefficients of inert small molecules.

**The model of small sticky molecules diffusing in polymer networks quantifies the effect of tube structures.**

Prediction of the diffusion coefficient of small molecules by hydrodynamic theories mainly depends on the empirical quantification of obstructions posed by polymer chains in the form of volume fraction, polymer and molecule sizes, etc (SI, Figure S6). We expect that the transport of small molecules could be effectively described by hydrodynamic theories since diffusion is inhibited by the presence of polymer chains and also inert molecules do follow the obstruction dependent transport mechanism as described in the previous section. However, the predictions of existing hydrodynamic models deviate from experimentally measured diffusion coefficient of small sticky molecules both in dilute and semi-dilute concentration regimes of the polymer (SI, Figure S6 and Table S3). To explain our observation of sticky molecules, we developed the transport model underscoring the local environment in the polymer solution instead of the coarse-grained approach in the case of an inert molecule.

To model our system, we describe the polymer network to consist of discrete compartments of specific length scales. In the case of sticky molecules, in the dilute regime, the important length scale is the correlation length and in the semi-dilute entangled regime, the relevant length scale is the tube size to properly describe their mobility in the polymer network (Figure 3A).

Our experimental data, combined with our prior work on molecular chemotaxis, led to hypothesize that molecules can interact with the polymer chains to result in enhanced mobility within the network. To provide a mechanistic understanding of the experimental data, we developed a *de novo* model of molecular diffusion driven by segmental friction and length scales within a complex, polymer network. First, we modeled the network as constrained virtual tubes or compartments of high molecular weight linear polymers (Figure 3A). The underlying physical depiction is the confinement of sticky molecules within tube compartments where it confronts both the fluctuating polymer chain and the local viscosity of the solvent (solution microviscosity). In dilute solutions, the small molecules are freely diffusing, akin to randomly moving around outside of these compartments, and are influenced by the local viscosity of the solution. In semi-dilute entangled regime, within the tube compartments, the sticky molecule can exhibit both free diffusion ($D_{local}$) and a weaker interaction-driven diffusion. The former component is governed by friction factor depending on the local viscosity of the environment and the latter component is dictated by the length scale of the compartment ($x(0)$) and a characteristic interaction time ($\tau_C$) such that the one-dimensional MSD of the molecule takes the following expression at small time $t$ (see Methods):

$$MSD_{@\ small\ t} \approx 2\ D_{local}\ t + \left(\frac{x(0)}{\tau_C/t}\right)^2 \quad (1)$$

where, $\tau_C \sim \frac{\gamma_s\, x_0}{k_h}$ is a lumped time parameter with underlying terms consisting of segmental friction factor of polymer chains ($\gamma_s$, N-s/m), decay length of sticky interaction ($x_0$, m) (SI, section S3) and a force constant of non-specific interactions ($k_h$, N). The details of this MSD model from the Langevin force balance is described in the methods section. Note that the MSD has frictional contributions from both local viscosity ($D_{local}$) and segmental friction ($\gamma_s$) inside tube compartments. Also, note that the $\tau_C$ parameter could be evaluated (SI, section S3 and Table S4) knowing $k_h$, $x_0$ and $\gamma_s$ values (eg. Rh6G case) or could be fitted to the experimental data (eg. Ficoll case).

For a sticky molecule diffusing in an entangled polymer solution, the random diffusion behavior is coupled to an interaction-driven component due to the formation of confinement tubes. It was defined by a characteristic length $x(0)$ in the model, where the molecule first starts to come under the influence of non-specific or sticky hydrophobic interaction (SI, section S3). Alternatively, $x(0)$ was the theoretical location of the molecule at time, $t = 0$ on the surface of the network tubes. We found that the hypothesis of $x(0) \sim a_t/2$ separates viscosity domains outside and inside tubes (Figure 3A) as well as couples both local viscosity-driven random diffusion and the weaker interaction-driven component inside tubes at $\tau_C$ timescale.

We can also check the limit on length scale in the case of sticky molecules diffusing in a dilute polymer solution. In the dilute regime, polymer chains are not yet overlapped and constrained structures are not present in the polymer microenvironment to confine the molecule and facilitate sticking. In other words, $a_t \to 0$ and as a result $D_x \to D_{local}$ which would be only driven by the microviscosity of the solution. This scenario takes place in case of a dilute solution of sticky molecules, for example, in the diffusivity of Rh6G. It should be noted that even in a dilute concentration regime, the mobility of sticky molecules cannot be explained by a $D_x = D_0\, e^{-N/N_c}$ type correlation which is obeyed by inert molecules.

Using our theory, we obtained good agreement between experimental measurements and theoretically estimated values in the case of Rh6G molecule at small time scales (Figure 3B). We modeled the data over three decades of MWs of polymers, spanning the dilute to semi-dilute regimes where polymer chains change from non-interacting random coils to entangled networks. For ease of comparison, we plotted MWs in log scale with base 10. Considering the presence of polymer chains in solution, we should expect that transport of both inert and sticky particles to get obstructed similarly. However, we observe contrasting mobility patterns between the molecules. According to our theory, this is because of the weak interaction-driven component of the mobility of sticky molecules, which nudges the molecule inside of the tube compartments where they could essentially diffuse similar to Brownian particles subject to the local viscosity. This mechanism, in case of the mobility of sticky molecules, causes the non-monotonic pattern which consists of an initial decrease in dilute regime followed by an increase and then decrease in semi-dilute entangled regime with the increase in polymer MWs (Figure 3B).

To show that our observation holds for other molecules of larger sizes, we performed measurements on a new molecule: TRITC-Ficoll of MW of $4 \times 10^4$ (abbreviated as Ficoll). We hypothesize that since Ficoll carries methylene groups it would interact with PEO chains and also due to its flexible nature[36], it is expected to localize into tube compartments easily. After performing diffusion coefficient measurements under the same conditions, we found that Ficoll indeed follows the mobility pattern of Rh6G (Figure 3C) similar to the previously described non-monotonic pattern of the diffusivity which initially decreases in dilute regime with an increase in MW followed by an increase which occurs due to chain overlap and formation of entanglements. Additionally, the modeling approach developed for Rh6G also applies to Ficoll (Figure 3C). Due to the significant size difference between Rh6G (diameter ~ 1.2 nm) and Ficoll (diameter ~ 3.0 nm), the absolute diffusion coefficients are different (Figures 3B and 3C), but their normalized values exhibit very similar patterns.

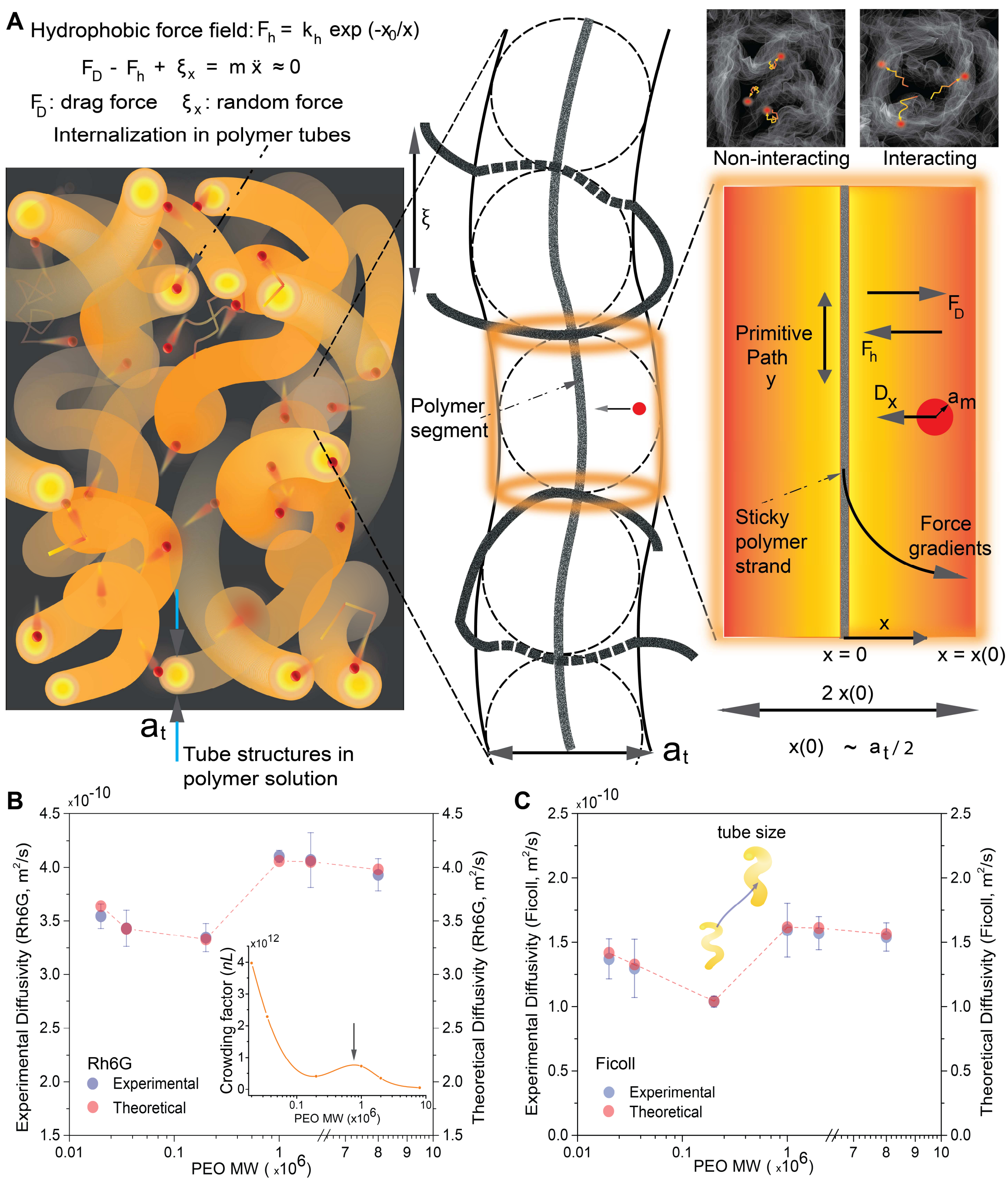

**Figure 3. Small sticky molecules diffuse within hypothetical tube compartments of polymer networks and subject to non-specific hydrophobic interactions, drag force, and Brownian random force as opposed to inert molecules, diffusion of which is simply dampened by the presence of polymer networks.** (A) The polymer segment is confined within a tube of diameter $a_t$ and has been assumed to contain interacting sites for the sticky molecule throughout its primitive path. The colored gradients in the vicinity of the primitive path depict hydrophobic force ($F_h$) gradients emanating from the polymer segment with a decay length $x_0$ and contact adhesion force $k_h$ as parameters of this non-specific interaction. A small molecule with a radius $a_m$ satisfying $2a_m \ll a_t$ shows deviation from standard SE diffusion principle within the network tubes and experience both local viscosity and segmental friction of polymer chains. Mechanistically, diffusion of sticky molecules would be facilitated by polymer networks and diffusion of inert molecules would be inhibited by polymer networks due to the absence of interaction. (B) Experimental measurements and theoretical modeling of diffusion coefficients of sticky Rh6G molecule in 0.5 wt% PEO solutions show good agreement between them. The inset shows the variation of crowding factor (*nL*) with MWs with a spline fit to estimated data. The arrow points at a local peak in diffusivity in the entangled regime of the polymer solution. (C) Experimental measurements and theoretical modeling of diffusion coefficients of sticky Ficoll molecule in 0.5 wt% PEO solutions show agreement between theory and experiments. The inset tube cartoons depict the increase in tube size in the entangled regime of the polymer solution. All error bars represent standard deviation from at least 5 independent measurements.

The interesting pattern of the mobility of sticky molecules can be explained in the light of spatial length scales of the network compartments in the polymer solution. Recovery of the diffusion coefficient of Rh6G, in PEO solutions with equal and greater than polymer MW of $2 \times 10^5$ could be explained in terms of tube compartment formation. In solutions of low MW of PEO, at the same concentration of 0.5 wt%, polymer chains do not overlap and confinements are absent. At 0.5 wt%, PEO chains would need to have MW around or greater than $\sim 2 \times 10^5$ (or 200K) to form entangled tubes. At this concentration, however, tube size is maximized in the vicinity of PEO MW of $10^6$ (SI, Figure S7). This is reflected in the crossover of the diffusion coefficient in the entangled regime as measured and shown in Figures 3B and 3C where the local maximum is observed at the same MW for both cases.

To generalize the length scale dependency of mobility and to explain mobility variation across MWs, we can define an effective crowding density of polymer chains by the crowding factor *nL*, where *n* is the number concentration of polymer chains and *L* is the length scale in the polymer medium. We find that the local maximum of the diffusion coefficient in the entangled regime depends on the crowding factor, irrespective of polymer concentration (SI, Figure S8). Additionally, the mobility variation across the range of MWs closely resembles the variation in crowding factor only if the molecule is interacting, irrespective of the type of the molecule. These observations directly suggest that the mobility of sticky small molecules is dictated by the structure of the polymer network and also suggests that different systems can adopt different magnitudes of $nL$ to modulate mobility according to the local conditions of crowding.

The observed local optima of diffusive mobility in entangled regime depend on the properties of polymer solutions and therefore, changes with polymer concentrations. We measured the Rh6G diffusion coefficient in 0.3 wt% PEO solutions and found that the observed local maxima in the diffusion coefficient in the semi-dilute regime shifted towards higher PEO MW of $2 \times 10^6$ (SI, Figure S9A) as compared to the case of 0.5 wt% PEO solutions. This upward shift in MW scale to cause local maxima in mobility is primarily driven by the fact that at 0.3 wt% PEO concentration, tube size gets maximized at around PEO MW of $2 \times 10^6$ (SI, Figure S7). A similar shift in mobility maxima is also demonstrated by sticky Ficoll molecules under the same conditions (SI, Figure S9B). Thus, generally speaking, the mobility of sticky molecules is affected by the relative length scale of the tube compartments at a particular concentration. The larger or more spacious the compartment, the higher is the diffusive mobility at a certain concentration of the polymer. However, at higher polymer concentrations, for example, in 1 wt% PEO solution, viscosity effects are also stronger and therefore, normalized diffusivity would be further dampened (SI, Figure S10).

To assess the effects of weak interactions in crowded conditions and to test an application of our theory, we evaluated diffusion coefficients of 5 nm gold nanoparticles (AuNP) in 0.5 wt% PEO solutions of increasing MWs using dynamic light scattering (DLS) measurements (SI, Figure S11). These 5 nm AuNP particles were coated with 5 kDa long PEG (polyethylene glycol) chains either terminated with methoxy groups (methylayed) or carboxyl groups (carboxylated). We hypothesized that methylated AuNPs could interact with PEO chains more effectively than carboxylated AuNPs and therefore, it might be possible to observe higher diffusivity of the former. Despite larger hydrodynamic sizes of these AuNPs and increased segmental friction from PEO chains, we observed a statistically significant level of difference between methylated AuNPs and carboxylated AuNPs with the former diffusing faster through PEO crowding (SI, Figure S11). Additionally, we confirmed that diffusivities of both methylated and carboxylated AuNPs correlate to crowding factor across the MWs tested with an optimized diffusion coefficient in the entangled regime as dictated by the crowding conditions (SI, Figure S11). Overall, we demonstrated a simple strategy to tune the transport of species by modulating weak interactions in the crowded environment.

**Unlike inert molecules, the MSD of a sticky molecule is not affected by the presence of polymers networks at small time scales.**

We modeled MSD of Rh6G molecule diffusing in PEO solutions following two approaches – (i) overdamped Langevin and (ii) one-dimensional Fokker-Planck (see Methods section and SI section S4 and also see SI Table S5 and Figure S12). These models are referred to as model_1 and model_2 in Figure 4A, respectively. We inverted the measured autocorrelation to estimate MSD at small time scales (0.4 ms) and then compared the MSDs between experimentally evaluated and theoretically modeled data (Figure 4A). We modeled MSD data at smaller time intervals using one-dimensional models due to the symmetric nature of the transport problem (see Methods). At small time scales, MSD profiles evaluated from both modeling and experiment are linear suggesting Brownian motion within the confocal volume irrespective of the interactions and confinements.

Underlying mechanistic insights could be obtained by contrasting MSDs between sticky and inert molecules under the same conditions. In the case of the sticky Rh6G molecule in the PEO solution, the MSD is larger than the inert 6-HEX molecule under similar conditions (Figure 4B). Hydrodynamically, the 6-HEX molecule is only 4% larger in size than Rh6G as measured in DI water. Therefore, we expect that the diffusion coefficient and corresponding mean square displacements to be ~ 4% larger for Rh6G. Yet the slope of the fitted line representing average MSD of Rh6G is around 64% higher compared to the 6-HEX case in PEO solution of MW of $2 \times 10^6$ (Figure 4B). Interestingly, we found that MSDs of Rh6G molecule in either DI water or in 0.5 wt% PEO solution of MW of $2 \times 10^6$ are very similar in values and molecular displacements of Rh6G are independent of the presence or absence of polymers as if the diffusing molecules are not hindered by polymer chains (Figure 4B and SI, Figure S9). Theoretically, it indicates the internalization of sticky molecules into the available space within tube compartments. However, at larger polymer concentrations such as 1.0 wt% PEO solutions, mobility is suppressed due to smaller compartment size and increased segmental friction from polymer chains, across the range of MWs (SI, Figure S10).

The absence of interaction is manifested in MSD profiles of inert molecules. In the case of inert 6-HEX molecules, a significant difference in MSDs was observed between DI water and 0.5 wt% PEO solution of MW of $2 \times 10^6$. However, unlike the sticky case, the MSD profile of this inert case was significantly dampened in 0.5 wt% PEO solution compared to DI water (Figure 4B). It suggests that inert molecules fail to internalize inside tube compartments at similar time scales and due to sensitivity towards the number of monomers and the polymer correlation length, they experience larger friction with increasing MWs and therefore, a greater extent of transport obstruction resulting in smaller MSD values. In other words, due to the absence of interactions, which suggests $k_h \rightarrow 0$, $\tau_C \rightarrow \infty$ and $MSD \approx 2\, D_0\, e^{-N/N_c} t$ following equation (1). Therefore, the lack of interaction would render normalized molecular mobility to be completely dependent on the physical properties of the polymer chain.

Although the MSD profiles appear to be almost linear at small time intervals for both sticky and inert small molecules (see Methods), the underlying distribution of the spatio-temporal displacements or probability distribution function might not be Gaussian [33]. We find that in cases of both Rh6G and 6-HEX molecules, the reduced kurtosis of autocorrelation function[37] at small time intervals are non-Gaussian (SI, Figure S12). Our modeling results also suggest that the Rh6G molecule, inside the tube, follows a probability distribution which is long-tailed instead of Gaussian (Figure 4B, inset). The Gaussian probability distribution cannot explain the MSD values subject to the constraints of our tube model. This observation supports recent developments in Brownian, but non-Gaussian dynamics [38–40]. Contrastingly, under similar conditions, 100 nm APSL particles which follow the macroviscosity of the solution, show Gaussian behavior with reduced kurtosis around zero (SI, Figures S3B and S12C)[37]. We observed a strongly decaying correlation of reduced kurtosis of measured autocorrelation function with molecular sizes of the diffusing species (SI, Figure S12C). This observation suggests a role of microviscosity in shaping the probability distributions of small molecules in crowded conditions. The underlying reason might be the different extent of medium heterogeneity probed by the particles of different sizes at a specific time interval of comparison[41].

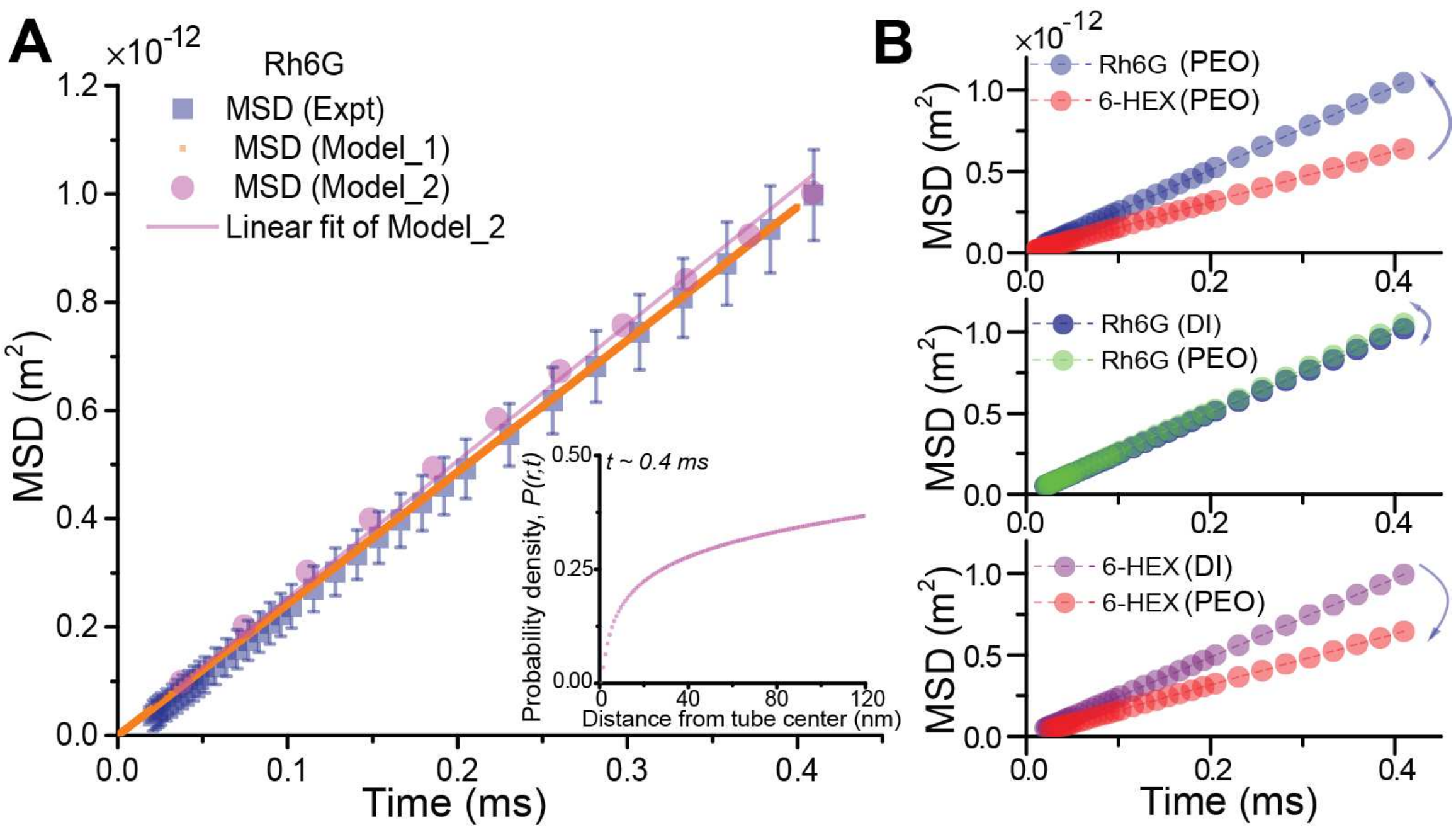

**Figure 4. Mean squared displacements (MSDs) of Rh6G, as well as 6-HEX molecules in three-dimensions show almost linear behavior at short time scales, but their MSDs are distinct based on the nature of interaction and presence of polymers in the solution.** (A) Estimation of MSD of Rh6G from experiments supports model predictions of isotropic three-dimensional MSDs as per the Langevin approach (Model_1) and the Fokker-Planck approach (Model_2 and linear fit of Model_2) under similar conditions (0.5% PEO of MW ~ $10^6$ or 1000K ) with error bars representing standard deviations of at least 5 independent measurements. The inset shows the modeled probability distribution function: $P(r,t)$ across the tube radius and at a time interval of ~ 0.4 ms. The apparent linear MSD is not driven by Gaussian distribution, but long-tailed non-Gaussian distribution within the tube. (B) Despite being similar in size, the estimated average MSD of Rh6G molecule is relatively higher than the 6-HEX molecule in the same PEO solution (MW ~ 2000K) under the same conditions. Average MSDs of Rh6G in DI water and 0.5% 2000K PEO solutions are equivalent and the molecule does not seem to feel the presence of polymer chains. THE average MSD of 6-HEX in 0.5% 2000K PEO solution is smaller than in DI water and the molecule is indeed inhibited by the presence of polymer chains. The average over 3 independent measurements of MSD is shown in each case of Figure 4B.

**Higher salt concentration enhances the mobility of sticky molecules and alters the diffusion patterns across all concentration regimes of polymer solutions.**

We can modulate the interaction by adding salt to the system, which presumably affects the hydrophobic interaction and not the electrostatic interaction as PEO chains are charge neutral. Such strengthening of hydrophobic interactions upon addition of higher salt concentrations is a common strategy that is used in chromatography and molecular chemotaxis [18]. We find that the presence of higher salt concentration in the PEO solution only altered the mobility pattern of sticky molecules significantly.

However, at lower salt concentrations, such as 1 mM KCl addition to PEO solution did not result in any noticeable difference from the DI water case, which further ruled out a major involvement of electrostatic screening effects.

In the absence of polymer crowding, there is no interaction and therefore, the addition of salt does not impart any difference in diffusive mobility. For example, diffusion coefficients of Ficoll in 1 mM and 10 mM KCl solutions were measured as 1.6 ± 0.1 × $10^{-10}$ m$^2$/s and 1.6 ± 0.2 × $10^{-10}$ m$^2$/s, respectively. Additionally, we also did not observe significant differences between diffusion coefficients in DI water and 1 mM KCl with the presence of PEO. Therefore, lower salt concentration does not boost interaction sufficient enough to impact the diffusion coefficient of the molecule in the presence of polymer crowding. This applies to both sticky and inert cases. However, the salt effect at higher concentrations (eg. 10 mM) and in presence of polymers, could alter mobility characteristics of the molecule if an interaction is also present. The possible underlying reason is the strengthening of hydrophobic interactions due to modification in decay length ($x_0$) upon an increase in salt concentration (see SI, section S3).

The diffusivity profile of sticky molecules also shows an interesting dependence at higher salt concentrations (eg. 10 mM) at all MWs in both dilute and semidilute regimes (Figure 5A and SI, Figure S13). At the same concentration of salt, both sticky and inert molecules are expected to get affected in similar ways. More strikingly, the diffusion coefficient pattern of only sticky molecules changes appreciably and inert molecules continue to exhibit the monotonic diffusion pattern as compared to lower salt concentrations (SI, Figure S13). The average increase of diffusion coefficient at 10 mM KCl concentration as compared to 1 mM KCl concentration varies over MWs and depends on the diffusing species, but generally remains higher in the dilute regime of polymer concentrations. As an example, in 0.5 wt% PEO solution of MW of $2 \times 10^4$, the average % of mobility increase in the cases of sticky Ficoll and Rh6G molecules were 16% and 11%, respectively, at the same conditions. Whereas, the average % increase of mobility of inert 6-HEX or RhB-PEG molecule was negligible under similar conditions (Figure 5A).

**A parametric model explains the mobility enhancement pattern of sticky molecules.**

The effect of higher salt concentrations on diffusion coefficients also reflects in MSD profiles of the molecules. At 0.5% PEO solution of MW of $2 \times 10^4$, the average MSD of sticky molecules increases due to the increase in salt concentrations. For example, in cases of Rh6G and Ficoll, there is a distinct increase in slopes of average MSDs while increasing salt concentration from 1 mM to 10 mM (Figure 5B). Contrarily, in the case of 6-HEX, under the same conditions, there is negligible change in the slope of MSD while increasing the salt concentration from 1 mM to 10 mM (Figure 5B).

Theoretically, the intricate relationships between salt effect and mobility at small time intervals can be explained by the characteristic interaction time parameter $\tau_C$. The underlying variables of $\tau_C$ are difficult to evaluate experimentally. However, we attempted to use $\tau_C$ as a phenomenological parameter to explain the experimental results. We found good agreement between experimentally measured and theoretically calculated diffusion coefficients of Ficoll in 0.5% PEO solutions of varying MWs at 10 mM KCl using our network compartment theory with $\tau_C$ as the sole fitting parameter (Figure 5C). In this modeling approach, we increased values of $\tau_C$ with polymer MWs spanning dilute, transition, and semi-dilute solutions of PEO. It is expected that with an increase in MW of polymer, the segmental friction ($\gamma_s$) would increase, which in turn would increase the value of $\tau_C$ parameter. Additionally, in the presence of salt, due to the strengthening of hydrophobic interactions, the decay length ($x_0$) might increase leading to further increase in $\tau_C$ parameter. Following these, we found a good agreement to experimental results by enhancing $\tau_C$ over the respective concentration regimes (Figure 5C). The variation of $\tau_C$ at different MW of PEO is shown in the inset of Figure 5C and follows an approximate sigmoid pattern. This phenomenological approach not only simplifies the complex interplay of many underlying variables of the system, but also quantitatively supports experimental results.

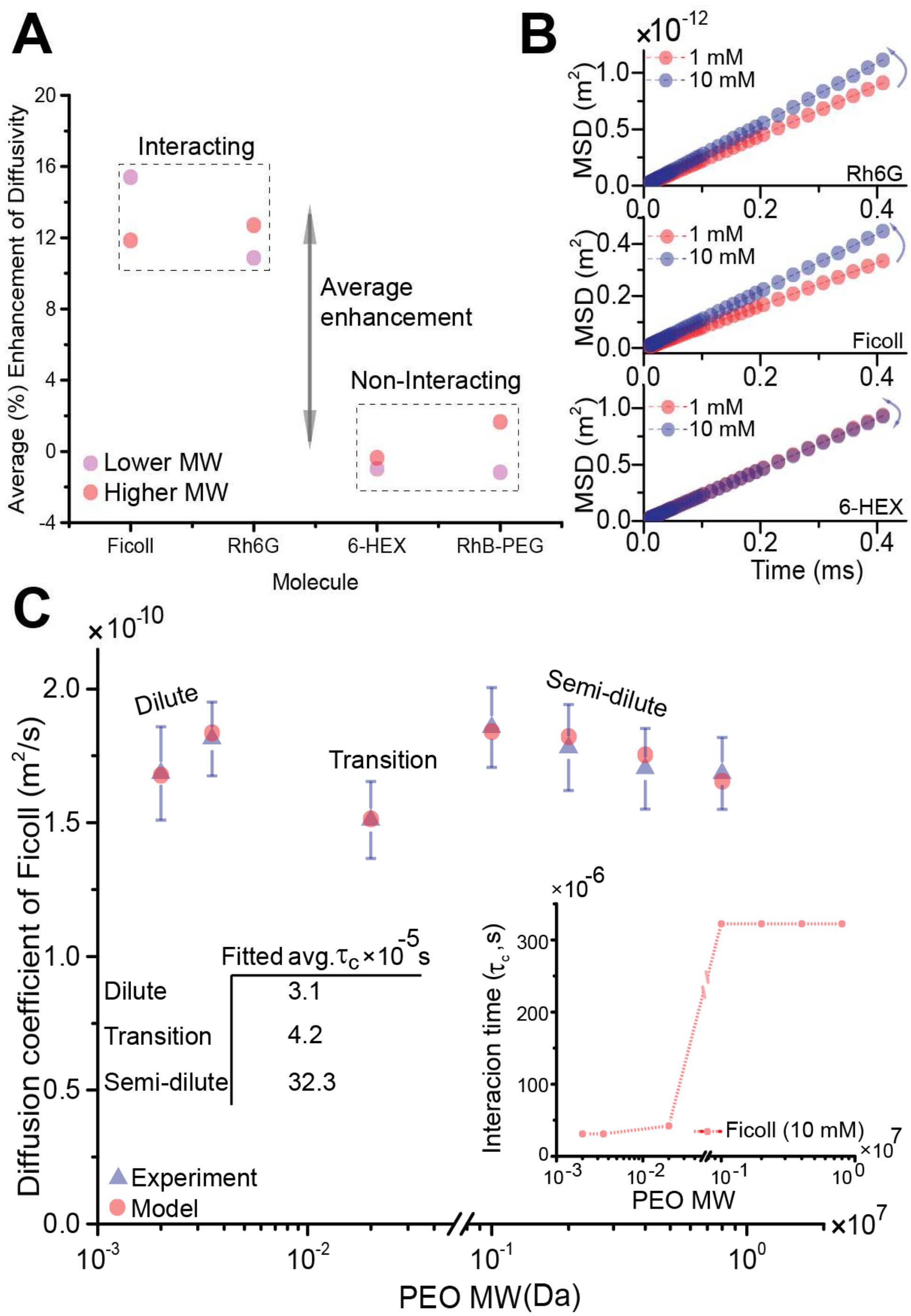

**Figure 5. Increasing salt concentration (1 mM to 10 mM) enhances diffusivity at different MWs (Da) of polymers, increases MSD of sticky molecules only, and changes the pattern of mobility variation across all concentration regimes of the polymer solution.** (A) The average increase of diffusion coefficient, from 1 mM to 10 mM KCl addition and in the case of sticky molecules (Ficoll, Rh6G) at either lower (20K) or higher (8000K) MW of PEO solution, is at least 10% higher than inert molecules (6-HEX, RhB-PEG). (B) MSDs of Rh6G and Ficoll in PEO solution of MW 20K show increase at 10 mM (blue) *w.r.t.* 1 mM (red) KCl concentration and the linear fitted lines of average MSDs at respective salt concentrations are discernible. However, MSD of 6-HEX in PEO of MW 20K and at 10 mM (blue) KCl is

not distinguishable and overlapping *w.r.t.* 1 mM (red) KCl concentration. (C) Modeling predictions of Ficoll diffusion coefficients in polymer network spanning all the concentration regimes of the polymer (dilute, transition, semi-dilute) support experimental measurements. The inset shows characteristic interaction time ($\tau_C$) of Ficoll with the increase of MWs of PEO. All error bars are standard deviations from at least 5 independent measurements.

**MT networks function at a specific crowding factor $nL$ which is evolutionary conserved.**

In the biopolymer networks, the observation of $nL$ dependent structural modulation similar to synthetic networks could be important as it might suggest enhanced transport of small sticky molecules despite the crowded medium. Intracellular systems are crowded partly because biological mechanisms necessitate proximity of materials to function and therefore, dissuade delocalization of machinery like enzymes, proteins, etc. However, to support biological reactions, small molecules also need access to proper channels of transport without much expense of energy. Here, we propose that the $nL$ factor might be one of such channels that could facilitate the passive transport of biologically relevant small molecules through intracellular crowding. In other words, we propose the hypothesis that there could a specific mesh size of the biopolymer network dictated by $nL$, which would correlate to the mobility of sticky small molecules and therefore, the metabolic state within the crowded vicinity.

Biopolymer networks are porous structures and are characterized by mesh size which is the average spacing between the filaments. For F-actin and MT networks, the tube size depends on the mesh size ($\xi_m$), which in turn depends only on the number of filaments per volume ($n$) and the contour length of the filaments (L) through $\xi_m \approx \sqrt{3/n\,L}$ [25].

The MT network is an essential structure in the cell that changes depending on cell cycle and cell maturation. Interestingly, through data analysis of published research, we find that normal, functioning MT systems operate within a particular $nL$ range, and in a diseased state, the $nL$ factor changes substantially, highlighting a potential link between altered mobility and diseased cell states (Supplementary Tables S6-S7). Our theoretical analysis does not state or confirm that $nL$ factor adopts a particular value only to modulate transport, but only presents a possibility that such an organization might correlate to transport modulation.

For various cellular MT systems (eg. metaphase spindles, flagellar axoneme, etc.) in the literature, we estimated the average $nL$ value of the system at steady-state. We found that the $nL$ value is mostly conserved in different species spanning across millions of years of evolution measured as per phylogenetic divergence time (Figure 6 and Supplementary Table S6). The average crowding density estimated was $nL \sim 3.75 \pm 0.52 \times 10^{14}\ \#/m^2$ (Figure 6).

To argue from a theoretical standpoint on the conservation of $nL$ value which could be correlated to the normal functioning of the cell and therefore, affecting the state of mobility in MT networks, we analyzed $nL$ data from diseased cells as a control case. The underlying assumption here is that the mobility of small sticky molecules in the intracellular matrix remains locally optimized at normal conditions of the surrounding milieu. An interesting observation from Figure 6 is that the metaphase spindle of HeLa cells operates further away from the conserved $nL$ value. HeLa cells are a cancer cell line, and due to its cancerous nature and altered metabolic state, there could be a divergence of $nL$ value from the conserved one. Excitingly, HeLa cells that were treated with 10 nM Taxol, a small molecule chemotherapeutic drug, had a network organization that was closer to the conserved $nL$ (SI, Table S7). To characterize the deviation of network structure, we defined the $nL$ dependent ratio parameter $r_{nL}$ which signifies the ratio between existing $nL$ of the system to the conserved $nL$ in normal MT networks. Specifically in HeLa cells, without any therapeutic intervention, $r_{nL} \sim 1.6$, and after 10 nM Taxol treatment, $r_{nL} \sim 1.1$, indicating recovery of normal functioning and intracellular mobility after correcting the $nL$ factor (SI, Table S7). These observations suggest a theoretical underpinning of the structural organization of intracellular spaces at a local level. Although not in the scope of the present research, an experimental understanding of the role of $nL$ factor and its bearing on local transport could assess the hypothesis on network organization and mobility correlation developed here.

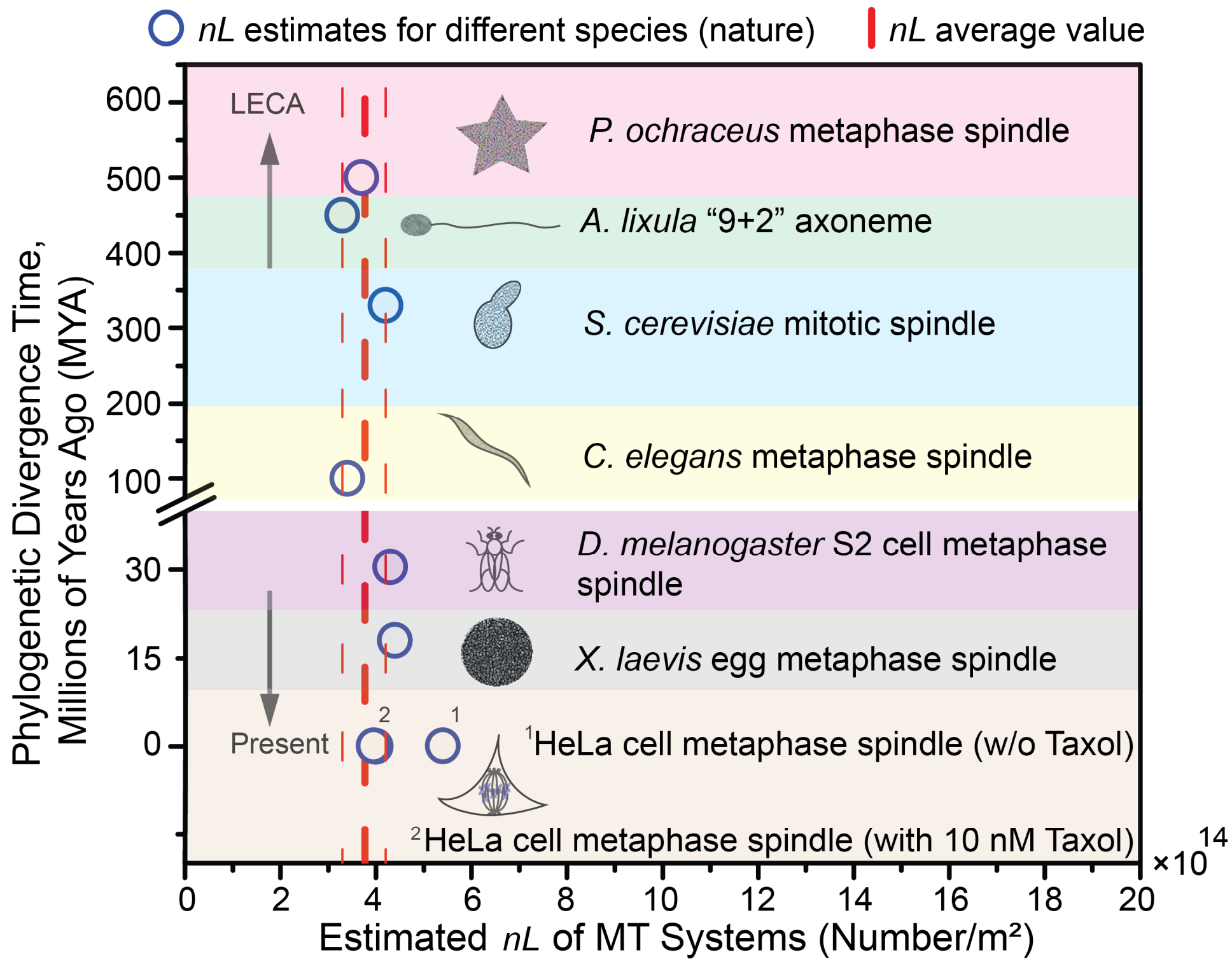

**Figure 6. Crowding factor $nL$ of rod-like polymers is conserved in different MT systems across phylogenetic divergence time (MYA) and species following the hypothesis of diffusive mobility modulation with the structural organization.** The diffusion-evolution plot shows that different organisms maintain an average $nL$ value (circle, blue) in the vicinity of $nL \approx 3.75 \pm 0.52 \times 10^{14}\ \#/m^2$ (average: bold dashed line, red; 95% confidence interval: normal dashed line, red). The y-axis represents an approximate time of divergence and evolution of that particular species from its closely related species in MYA from the present time (see SI, Table S7). The upward-pointing arrow in the figure could be stretched to ~ 1000 MYA till the diversification of the last eukaryotic common ancestor (LECA), since when the axoneme structure has been in use. Different MT systems, including cells of sea-star (*Pisaster ochraceus*), sea urchin (*Arbacia lixula*), yeast (*Saccharomyces cerevisiae*), nematode worm (*Caenorhabditis elegans*), fruit fly (*Drosophila melanogaster*), frog (*Xenopus laevis*) and human (HeLa cell) - all converge on the same mean-field steady-state value of $nL$.

**Discussions:**

Interaction-driven diffusion phenomena in equilibrium systems have come to scientific purview very recently [42–44]. In the presence of concentration gradients, diffusiophoresis[45], cross-diffusion [18], binding-driven phoresis [46], or hydrodynamic volume exclusion [5] was proposed to facilitate the transport of molecules and colloids. In homogeneous macromolecular crowding, the entanglement of polymer chains and non-specific, weak interactions facilitate the transport of the 'sticky' class of molecules, but impede transport of the 'inert' class of molecules. The proposed passive transport mechanism enables the localization of sticky small molecules into the compartments of network meshes and tubes. Unlike inert molecules, the mobility of sticky molecules could be enhanced by changing the local physicochemical environment of the macromolecular crowding. Overall, the short-range non-specific interactions could enable global change in both magnitude and pattern of the mobility of small molecules, as observed in case of molecular chemotaxis in concentration gradients (SI, Figure S1).

Here, we show that interaction is an essential component of mobility depending on the nature of the small molecule, irrespective of its size, and the mobility also depends on the properties of the polymer network. For the inert case, the mobility varies macroscopically with the number of monomer units in the polymer chain and a characteristic constant for the system $N_c$ which could be correlated to crowding concentration or correlation length of the system. This simplifies the model prediction as opposed to tracking multiple system-specific parameters used in other models [47–49]. We also found that the normalized diffusion coefficient for such inert small molecules and macromolecules does not depend on the molecule itself, but on the properties of the polymer network.

On the contrary, sticky molecules diffuse in polymer solutions following a fundamentally different mechanism compared to inert molecules. Sticky molecules can localize within the hypothetical compartments made by network tube structures and interact with the polymer segments which would alter their mobility. We found that these weak interactions could be modulated by (i) changing the surface groups of the species, (ii) changing the polymer MWs, and (iii) the salt concentration of the medium. These molecules cannot feel the presence of polymers at certain concentrations and MWs around $\tau_C$ timescale which is of the order of our time interval of measurements. This observation not only suggests that larger free space is available to the molecule to move within network compartments and the importance of crowding factor $nL$ in tuning the mobility, but also indicates the possibility of crawling by the diffusing molecules along the polymer chains[50] facilitated by interactions. In DI water or at low salt concentrations, length scale attributes dominate, and at higher salt concentrations interaction dominates and $\tau_C$ becomes important. Overall, $\tau_C$ could be used as a knob to modulate the mobility of sticky molecules in polymer solutions. In summary, small molecules could be classified into two classes – sticky and inert, through diffusion behavior in polymer networks.

Similar to the discussed MT systems where *nL* factor seems evolutionarily conserved for functionality, network structure and mesh size of LLPS droplets also might be correlated to transport modulation through the *nL* factor. It has been observed that LLPS structures facilitate entries of certain client molecules that target the milieu and restrict others [26]. This is analogous to selective transport and interactions of sticky and inert molecules through proper mesh size within synthetic polymer networks. Additionally, the reported mesh size of the order of ~ 10 nm within P-granules [51] could facilitate the transport of interacting small molecules such as ATP, smaller proteins, etc. Additionally, the driving constituents of LLPS, the so-called intrinsically disordered proteins, diffuse faster in crowded environment [52] and could be affected by weak interactions.

More broadly, the discussed experiments and theories would help to establish Brownian, yet non-Gaussian diffusion phenomena observed in the context of small molecules experiencing microviscosity to quantitatively explain transport and organization of intracellular systems where both Brownian and anomalous diffusion have been observed [53]. The experimental framework could be used to find intelligent molecules and design intelligent systems for applications relevant to sensing and diagnostics. Additionally, the demonstrated mobility optimization guideline could be useful in developing faster and safer rechargeable batteries with improved design principles for electrode coatings and other polymeric components.

## Methods

**The equations and derivation of the diffusion coefficient of a sticky molecule:**

When the polymer concentration ($c$) in the medium exceeds the critical concentration ($c^*$), the translational isolation of the linear polymer chain becomes stronger, and it becomes laterally confined into a hypothetical tube structure of diameter $a_t$ which is constrained by surrounding chains [54]. Any small molecule satisfying $2a_m/\ a_t \ll 1$ would be completely confined within the topologically constrained tubes with local viscosity dictating its Brownian motion. In this stable tube conformation (segmental relaxation time ≪ diffusion time ≪ Rouse relaxation time, SI, Tables S1 and S5), hydrophobic interactions between the polymer chain and the small molecule would internalize the molecule into the free tube space and the molecule would diffuse as if no barrier to diffusion is present. Please note that the actual polymer chain is oriented along the primitive path following the centerline of the tube (Figure 3A). Here, part of the polymer chain is designated as the segment as shown in Figure 3A. Since curvilinear diffusion of this polymer segment (y-direction) is negligible compared to the diffusion of the small molecule, which is perpendicular *w.r.t.* the direction of the curvilinear diffusion or primitive path at the time interval considered, we simplified the model along the x-direction only (Figure 3A).

**Langevin approach (Model_1):**

According to this one-dimensional diffusion model subject to the interaction force $F_h$, a generalized Langevin equation in the x-direction could be expressed:

$$m\ddot{x} = F_D + F_h(x) + \zeta_x(t) \tag{2}$$

Where the drag force is denoted as $F_D = \gamma\dot{x}$, attractive hydrophobic force is represented by $F_h$ and $\zeta_x$ is the fluctuating stochastic force imparted by solvent molecules, $m$ is the molecular mass, $\ddot{x}$ is the molecular acceleration, γ is the friction factor, $\dot{x}$ is the velocity of the molecule. The inertial effects in eqn (2) can be neglected even in μs – ms time scale, which led us to simplify eqn (2) into the overdamped Langevin equation as follows:

$$\gamma \frac{dx}{dt} - F_h(x) + \zeta_x(t) \approx 0 \tag{3}$$

Notice that the directions of $F_D$ and $F_h$ are opposite in our system. The hydrophobic force is known to follow an exponential decay profile with a characteristic decay length ($x_0 \sim 1.0$ nm, see SI, section S3) and a pre-factor or adhesion force ($k_h$) given by: $F_h(x) = k_h\, e^{-x/x_0}$, where $k_h$ depends on the radius of the diffusing molecule ($a_m$) and a spring force constant $C$ (~ 0.1 N/m) such that $k_h = a_m C$ [55]. To incorporate the non-specific, sticky interaction with an equilibrium association constant $K_a$, $k_h$ could be expressed in a slightly different form: $k_h = \frac{1}{x_0} K_B T\, ln(c^0 K_a)$, where $c^0 = 1$ M, $K_a$ is expressed in $M^{-1}$ (see SI, section S3).

Using integration factor $e^{\frac{k_h t}{\gamma x_0}}$, we find the integrated form of eqn (2) as follows with an assumption of $e^{-x/x_0} \sim 1 - x/x_0$ :

$$x = \int_0^t \frac{1}{\gamma} \left(k_h - \zeta_x(\tau)\right) e^{\frac{k_h}{\gamma x_0}(\tau - t)} d\tau + c\ e^{-\frac{k_h t}{\gamma x_0}}$$

Taking time average over the traversed displacement $x$ and using the initial condition $\langle x \rangle|_{t\to 0} \sim x(0)$ which signifies the entrance to the tube compartment and then averaging over Gaussian random noise: $\langle \zeta_x(\tau) \rangle = 0$, we can simplify the above equation into the following form:

$$\langle x \rangle = x_0 \left(1 - e^{-\frac{k_h t}{\gamma x_0}}\right) + x(0)\, e^{-\frac{k_h t}{\gamma x_0}} \tag{4}$$

To calculate square displacement, we estimate $x^2$ with time variables $\tau$ and $\tau'$ such that-

$$x^2 = \frac{1}{\gamma^2} \int_0^t \left(k_h - \zeta_x(\tau)\right) e^{\frac{k_h}{\gamma x_0}(\tau - t)} d\tau \int_0^t \left(k_h - \zeta_x(\tau')\right) e^{\frac{k_h}{\gamma x_0}(\tau' - t)} d\tau'$$
$$+ \frac{C}{\gamma} \int_0^t \left(k_h - \zeta_x(\tau)\right) e^{\frac{k_h}{\gamma x_0}(\tau - 2t)} d\tau$$
$$+ \frac{C}{\gamma} \int_0^t \left(k_h - \zeta_x(\tau')\right) e^{\frac{k_h}{\gamma x_0}(\tau' - 2t)} d\tau' + C^2\ e^{-\frac{2 k_h t}{\gamma x_0}}$$

Where, $C = x(0)$. Following the above expression, $\langle x^2 \rangle$ can be simplified with the application of random fluctuations and autocorrelation in the form of Dirac delta function with an amplitude of diffusion coefficient $D$ and with the following form: $\langle \zeta_x(\tau)\, \zeta_x(\tau') \rangle = 2\, D\, \gamma^2 \delta(\tau - \tau')$, along with $\langle \zeta_x(\tau) \rangle = 0$. The form of $\langle x^2 \rangle$ is as follows-

$$\langle x^2(t) \rangle = x_0^2 \left(e^{-\frac{2 k_h t}{\gamma x_0}} - 2\, e^{-\frac{k_h t}{\gamma x_0}} + 1\right) + D \left(\frac{\gamma\, x_0}{k_h}\right)\left(1 - e^{-\frac{2 k_h t}{\gamma x_0}}\right) + 2\, C\, x_0 \left(e^{-\frac{k_h t}{\gamma x_0}} - e^{-\frac{2 k_h t}{\gamma x_0}}\right) +$$
$$C^2 e^{-\frac{2 k_h t}{\gamma x_0}} \tag{5}$$

The mean square displacement (MSD ~ $\langle (x(t) - x(0))^2 \rangle$) with initial position $x(0)$ can be further simplified in the following form:

$$MSD = (x(0) - x_0)^2 \left(e^{-\frac{k_h t}{\gamma x_0}} - 1\right)^2 + D \left(\frac{\gamma\, x_0}{k_h}\right)\left(1 - e^{-\frac{2 k_h t}{\gamma x_0}}\right) \tag{6}$$

We are interested in assessing the fate of the molecule at smaller time scales, rather than at longer times when the molecule might move subdiffusively, cease motion due to stronger association with the polymer, or could execute a one-dimensional movement along the polymer chain. We expanded the exponential $e^{-at} \approx 1 - at + \frac{a^2t^2}{2}$ , neglected $(\frac{k_h t}{\gamma x_0})^2$ compared to $\frac{k_h t}{\gamma x_0}$ and assumed $x(0) \gg x_0$ (under the experimental conditions: $x(0) \sim$ 100 nm and $x_0 \sim$ 1 nm) to arrive at the following simplified form from eqn (6):

$$MSD_{@\,small\,t} \approx 2\,D\,t + \left(\frac{k_h}{\gamma\,x_0}\right)^2 x^2(0)\,t^2 \qquad (7)$$

The characteristic interaction time $\tau_C \sim \frac{\gamma\,x_0}{k_h}$ can be used as a parameter to simplify the MSD at small times as follows:

$$MSD_{@\,small\,t} \approx 2\,D\,t + \left(\frac{x(0)}{\tau_C/t}\right)^2 \qquad (8)$$

In the above equation, at short time intervals *t,* which is of the order of $\tau_C$, it is implied that there would be a boosting component in MSD over Brownian one ($2Dt$) which is dictated by local viscosity. This local viscosity or microviscosity-driven diffusion component is denoted as $D_{local}$ in the main text. It is important to note that we assumed isotropic MSD in all dimensions and used 3× multiplication factor over one-dimensional MSD evaluated from theoretical calculations (SI, Table S5) and found agreement with experimental data. Therefore, the theoretical consideration of the one-dimensional movement of species within tubes is substantiated. We also found that polymer chain dynamics have a weaker contribution to MSD at small time scales and we neglected it in our existing model (SI, Table S5).

**Fokker-Planck approach (Model_2):**

We used one-dimensional Fokker-Planck equation in cylindrical co-ordinates due to radial symmetry of diffusive mobility in the tube compartments as per the following form:

$\frac{\partial P}{\partial t} = D\,\frac{1}{r}\,\frac{\partial}{\partial r}(r\,\frac{\partial P}{\partial r}) + \frac{1}{\gamma}\,\frac{\partial}{\partial r}(P\,\frac{\partial V}{\partial r})$ , where the potential $V$ is due to hydrophobic forces: $F_h(r) = -\,\frac{\partial V}{\partial r}$ and $P(r,t)$ is the probability density function of the molecule in the tubes.

We evaluated the Fokker-Planck equation in both one dimensional rectangular (linear: along *x*) and cylindrical coordinates (radial: along $r$) and found ~10% variation in calculated MSDs between the two methods. We only explained the cylindrical form of the equation here.

Replacing $F_h(r) = k_h\, e^{-r/x_0}$ and rearranging terms, we get:

$$\frac{\partial P}{\partial t} = D\,\frac{\partial^2 P}{\partial r^2} + \left(\frac{D}{r} - \frac{k_h\, e^{-r/x_0}}{\gamma}\right)\frac{\partial P}{\partial r} + \frac{k_h}{\gamma\, x_0}\, e^{-r/x_0}\, P$$

By defining dimensionless variables of time: $\tau = t/\tau_C$ and of position: $\eta = r/x_0$ , we simplify the above equation to the following form:

$$\frac{\partial P}{\partial \tau} = k'\,\frac{\partial^2 P}{\partial \eta^2} - \left(e^{-\eta} - \frac{k'}{\eta}\right)\frac{\partial P}{\partial \eta} + e^{-\eta}\, P \qquad (9)$$

where constant: $k' = \frac{D\,\tau_C}{x_0^2}$ . The initial and boundary conditions on the dimensionless probability density $P(\eta, \tau)$ are the following:

1. Initial condition: $P(\eta, 0) = P_{ini}\,\delta(\eta\, x_0)$ which originates from the fact that the initial probability of finding the dye molecule in the polymer solution, *i.e.* $P_{ini} = \frac{C_{dye}}{C_{dye} + C_{pol} + C_{water}}$ , where $C_{dye}$, $C_{pol}$ and $C_{water}$ are initial concentrations of dye, polymer, and water molecules in the solution. $\delta(\eta\, x_0)$ is the Dirac delta function at position $r$.

2. Boundary conditions: (i) At the compartment boundary ($\eta \to \eta_{tube}$), we used a power-law diffusivity profile to generate long-tailed distribution to explain Brownian yet non-Gaussian diffusion[33] with exponent $\alpha$ ($1 < \alpha < 2$) and of the following form:

$$P(\eta, \tau)|_{\eta \to \eta_{tube}} = \frac{(a_t/2)^2}{(\eta_{tube}\, x_0)^{2\alpha+1}}$$

Good agreement to the experimental data was obtained with $\alpha \sim 1.15$.

(ii) On the polymer strand ( $r \to a_m$ or $\eta \to \eta_m$ ), we can derive the probability density as follows: $P(\eta, \tau)|_{\eta \to \eta_m} = P_0/a_m$ (SI, section S4).

In calculating the diffusion coefficients, we analyzed the autocorrelation data for small time intervals (~ 0.4 ms). We found that both types of small molecules, irrespective of their interaction with polymer network, exhibits non-Gaussian diffusion traits at small time intervals (SI, Figure S12 and section S5.1).

Therefore, accurate estimation of diffusion coefficient, as well as mean squared displacements (MSD) using FCS relies on the condition that the time interval of displacement under consideration be much smaller than the time required for the diffusing molecule to traverse the confocal beam waist (SI, section S5.2) [56]. Additionally, the constrain of small time intervals enables us to invert the autocorrelation to more accurately quantify MSD irrespective of the type of associated probability function [56]. The other reason behind the adoption of small time scale analysis is the characteristic residence time of molecules ($\tau_{residence}$) within tubes, which is of the same order of measurement time and thus provides the time scale necessary for the small molecule to explore the local environment. For example, in case of a particle diffusing at $D{\sim}10^{-10}\,m^2/s$ within tube size of $a_t{\sim}200\,nm$, the time scale of traversing the length scale: $\tau_{residence} \sim\ {a_t}^2/D\ \approx 4\,ms$.

**Data availability**

The basic algorithm and associated correlations for theoretical calculations of diffusion coefficients are shown in SI, section S6. Matlab codes used for simulating the theoretical model is available online at the GitHub repository with detailed instructions: https://github.com/rajarshiche/Anomalous-diffusion-in-polymer-crowding

**Additional Information**

**Associated Content**

Supplementary Information accompanies this paper and is available free of charge.

Additional information on the comparison of chemotaxis between sticky and inert molecules in microfluidic system, model of non-interacting small molecules, hydrophobic force expression in terms of a pre-factor, derivation of boundary condition on the polymer strand at equilibrium, FCS data analysis, and basic calculations and algorithm for diffusion coefficient estimation in PEO systems are in the supplementary.

**Author Information**

Corresponding Author: Rajarshi Guha: rajarshiche@gmail.com

**Notes**

The authors declare no competing interests.

**Acknowledgments**

The work was supported by Penn State MRSEC funded by the National Science Foundation (DMR-1420620). We acknowledge the help of Farzad Mohajerani in preparing solutions for measurements and Prof. Jacinta C. Conrad for helping to review the manuscript.

# SUPPORTING INFORMATION

# Entanglement and weak interaction driven mobility of small molecules in polymer networks

Rajarshi Guha[a,*] Subhadip Ghosh[c], Darrell Velegol[a], Peter J. Butler[b], Ayusman Sen[c], Jennifer L. Ross[d]

[a]Department of Chemical Engineering, Pennsylvania State University,

University Park, Pennsylvania, 16802, USA

[b]Department of Biomedical Engineering, Pennsylvania State University,

University Park, Pennsylvania, 16802, USA

[c]Department of Chemistry, Pennsylvania State University,

University Park, Pennsylvania, 16802, USA

[d]Department of Physics, Syracuse University

Syracuse, NY 13244

[*]Correspondence should be addressed to R.G. (rajarshiche@gmail.com)

## CONTENTS

Figures S1-S13

Tables S1-S7

**S1. Comparison of chemotaxis between sticky and inert molecules in microfluidic system:**

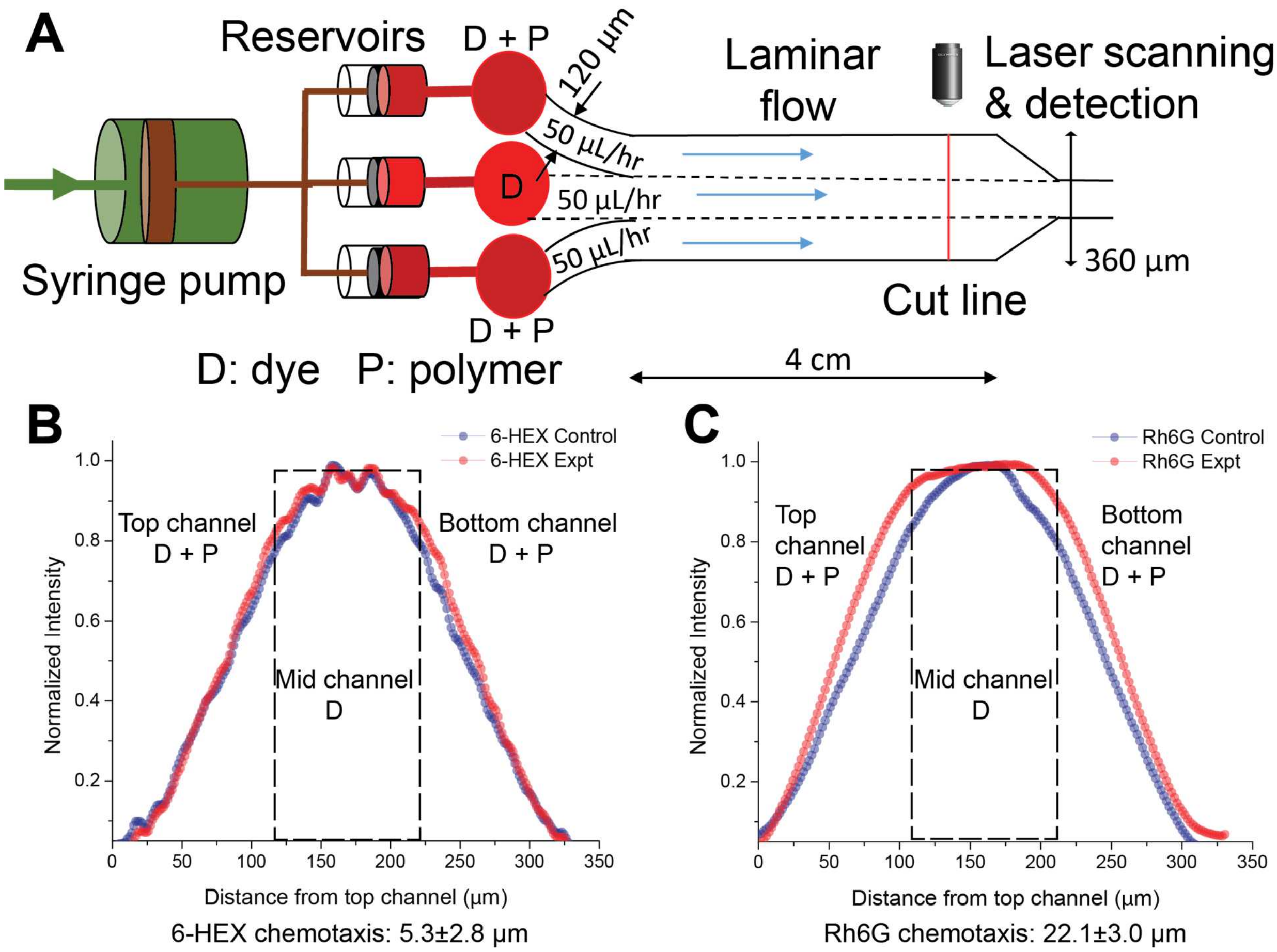

**Figure S1. Chemotaxis of Rh6G molecule in gradients of 0.5 wt% PEO solution of MW of $2 \times 10^5$ is 4-fold larger than 6-HEX chemotaxis under the same conditions in a 3-channel microfluidic setup.** (A) Schematic of the microfluidic experiment as previously described by Guha *et al.*[1]. All 3 channels flow the same concentrations of dye (D) at 50 µL/hr using a syringe pump and the flanking channels (top and bottom) also flow PEO (P). The observation of dye intensity was recorded near the end of the channels at a certain cut line for both control and chemotaxis experiments. (B) Spreading of 6-HEX molecules from the mid channel to flanking channels is not significant in gradients of PEO *w.r.t.* control case as measured from normalized intensity distribution. (C) Spreading of Rh6G molecules from the mid channel to flanking channels is quite significant in gradients of PEO *w.r.t.* control case as measured from normalized intensity distribution. The average estimated chemotaxis of 6-HEX molecules was 5.3±3.8 µm and the average chemotaxis of Rh6G molecules was 22.1±3.0 µm according to the method described by Guha *et al.*[1]. All values reported are averaged over at least 3 independent measurements.

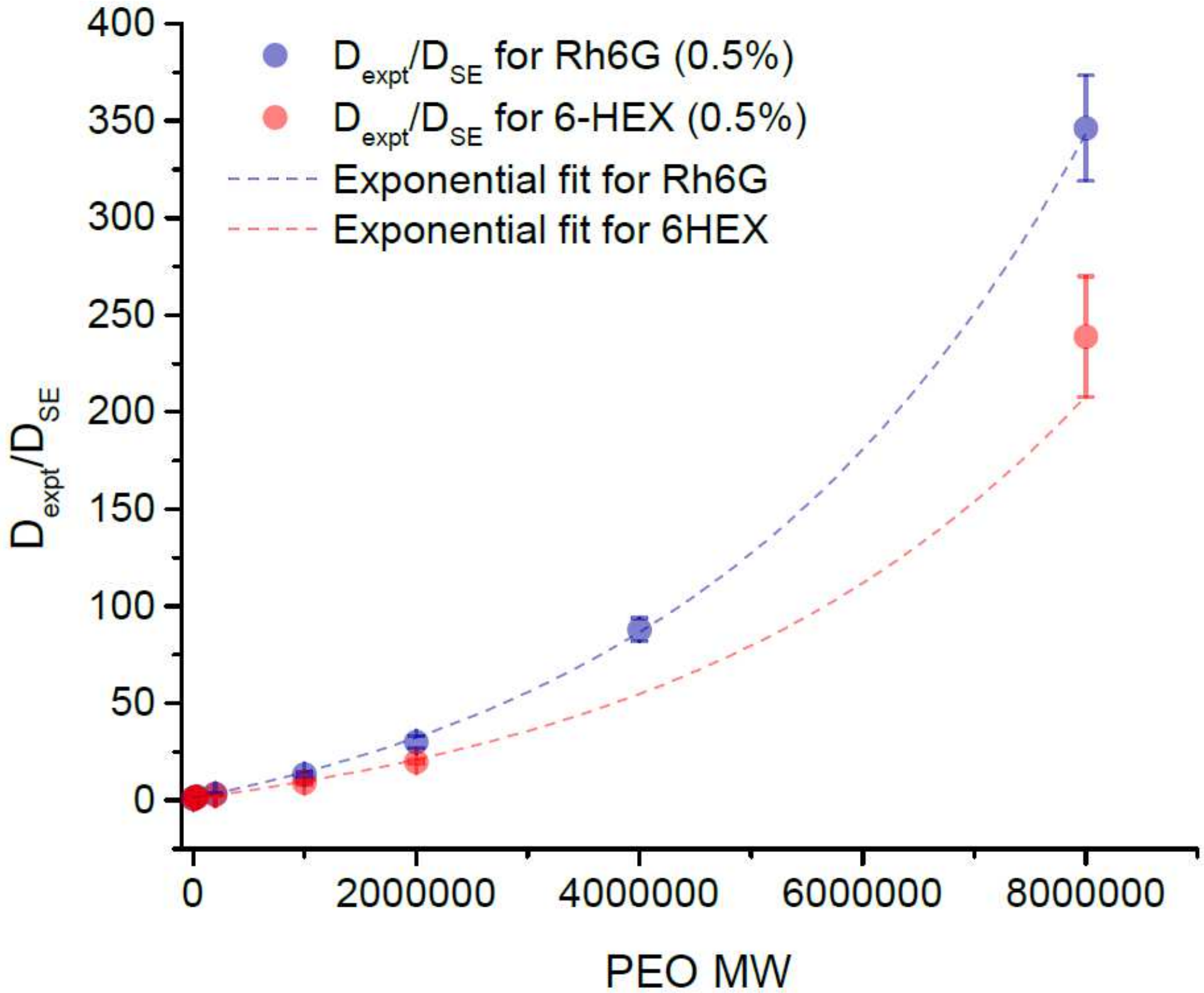

**Figure S2. The exponential increase of experimentally measured diffusion coefficient over the Stokes-Einstein diffusion coefficient ($D_{expt}/D_{SE}$) based on solution macroviscosity in case of Rh6G (blue) and 6-HEX (red) molecules diffusing in 0.5 wt% PEO solutions of different MWs.** $D_{expt}$ represents the average value over at least 5 independent measurements. Notice that the ratio diverges more with MWs in the case of Rh6G than 6-HEX.

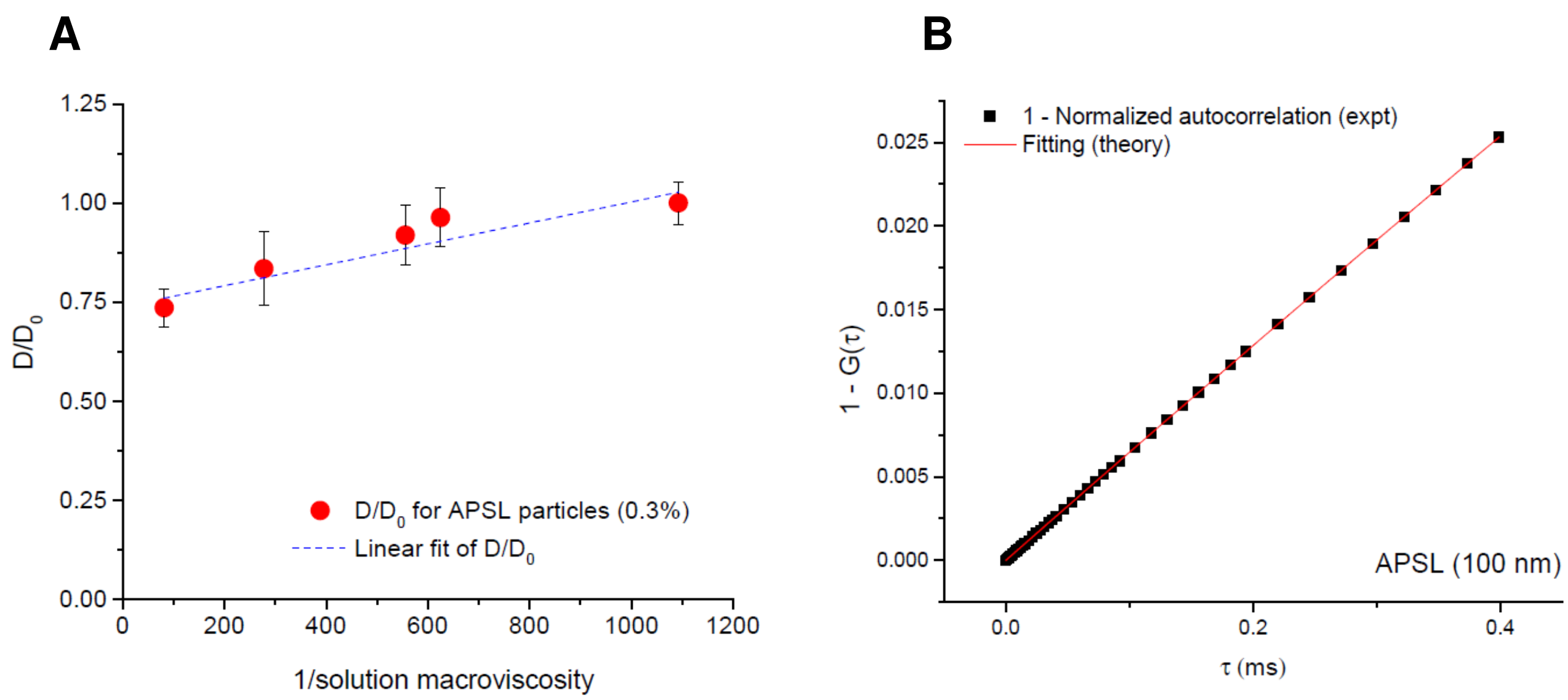

**Figure S3. Diffusion of 100 nm APSL particles in 0.3 wt% PEO solution follows the Stokes-Einstein theory.** (A) Linear plot of normalized diffusion coefficient ($D/D_0$) of 100 nm APSL particles in 0.3 wt% PEO solution with the inverse of solution macroviscosity (or bulk viscosity) follows the Stokes-Einstein diffusion pattern of colloidal particles. $D$ is the experimentally measured diffusion coefficient at any polymer MW (*i.e.* $D_{expt}$) and $D_0$ is the measured diffusion coefficient in DI water. The variation of solution macroviscosity is solely due to the increase of MWs of PEO chains. (B) After fitting normalized autocorrelation, $G(\tau)$ in the form: 1-$G(\tau) = -c_2 t^{\alpha} - c_4 t^{2\alpha}$ within the measurement time scale of our experiments (0.4 ms), we found the reduced kurtosis $K \sim -\ 0.03$ which corresponds to the Gaussian nature of APSL displacement [2]. All error bars represent standard deviations from at least 5 independent measurements.

**S2. Model of non-interacting, inert small molecules:**

**1-dimensional obstruction model**: Within a certain time interval and in absence of polymer molecules, the unobstructed transport length $l$ is available to a diffusing molecule to traverse from point A to point B in the solution as shown below (Schematic S1). However, in the presence of obstructions posed by monomeric blobs of the polymer chain (represented by blue spheres in the below schematic), the available transport path length decreases. With uniformly distributed N monomer blobs along the previously available transport length $l$, the obstructed transport length between two blobs available to the molecule is denoted as $\Delta l$. We now implement two simple criteria inspired from 1st order kinetics of chemical reaction into this one dimensional obstructed transport case:

(i) The longer the unobstructed transport length, the longer is the obstructed path length at the same degree of obstruction and within the same time interval. In other words, $\frac{\Delta l}{\Delta N} \propto l$ (1)

(ii) Since the obstructed transport path length decreases when the degree of obstruction increases, the corresponding slope should have a negative sign. In other words, $\frac{\Delta l}{\Delta N} \approx -\frac{dl}{dN}$ (2)

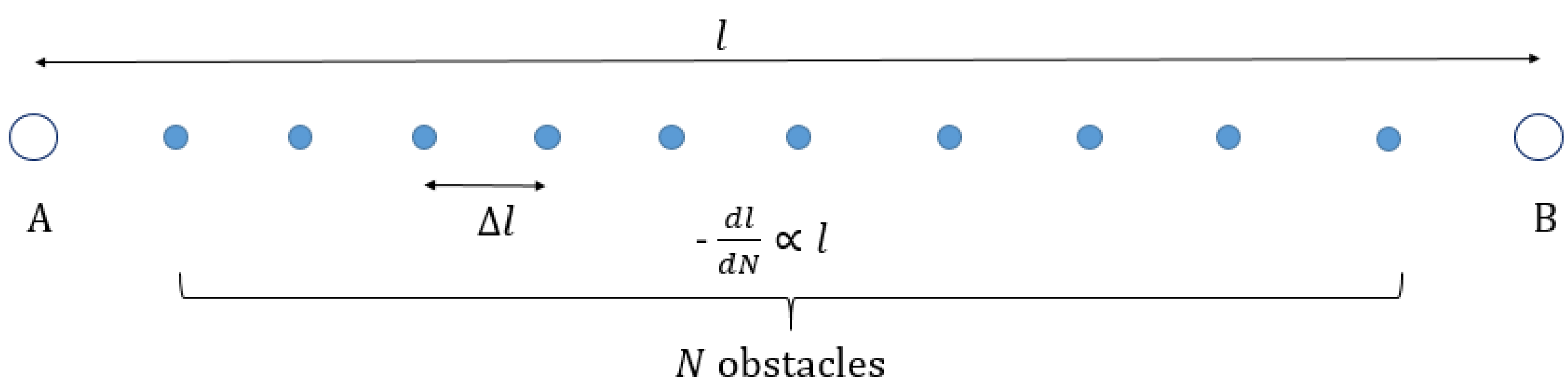

**Schematic S1. 1-dimensional obstruction model of non-interacting molecules diffusing in polymer solution of N monomer blobs.**

Additionally, as the monomer blobs are fluctuating in the solution, the diffusing molecule, in terminal cases, might not encounter any of them (0 obstacles) or all of them ($\Delta N$). Considering the average scenario of the degree of obstruction, i.e. $\Delta N/2$, we can express the proportional relationship between obstructed and unobstructed transport lengths as the following $\frac{\Delta l}{\Delta N/2} \propto l$. After inserting the sign term for slope as per criterion (ii) mentioned above and rearranging, we can further write $dl/dN \propto -\frac{1}{2} l \Rightarrow dl/l = -\frac{1}{2N_c} dN$. Where, $1/N_c$ is a constant. After integrating the differential equation between transport lengths $l$ and $l_0$ and the number of monomer blobs *N* and 0, we get the following equation:

$$\ln\left(\frac{l}{l_0}\right) = -\frac{1}{2N_c} N \quad (3)$$

With the diffusion coefficient $D_x$ as the transport coefficient in the presence of polymer and $D_0$ in absence of it, we can express the path lengths at time *t* in one dimension as follows: $l \sim \sqrt{2\, D_x\, t}$ and $l_0 \sim \sqrt{2\, D_0\, t}$. Inserting these terms in the above equation gives-

$\ln\left(\frac{\sqrt{2\, D_x\, t}}{\sqrt{2\, D_0\, t}}\right) = -\frac{1}{2N_c} N \Rightarrow \ln\left(\frac{D_x}{D_0}\right) = -\left(\frac{N}{N_c}\right)$. Therefore, the final expression of the diffusion coefficient of a non-interacting molecule obstructed by N monomer blobs of the polymer chain can be written as follows:

$$D_x = D_0\, e^{-N/N_c} \quad (4)$$

Please note that this model assumes a certain number of $N$ in the solution or a particular concentration of the polymer.

**Table S1 Estimated time scales in 0.5 wt% PEO solution of MW ~ $10^6$ suggest stable mesh conformation around the diffusion time scale**

| Segment relaxation time ($\tau_s$, sec) | Entanglement time ($\tau_e$, sec) | Diffusion time of 6-HEX in polymer mesh ($\tau_D \sim \xi^2/D_{6-HEX}$, sec) | Rouse relaxation time ($\tau_R$, sec) | Tube disengagement time ($\tau_d$, sec) |
|---|---|---|---|---|
| $1.1\times10^{-11}$ | $2.8\times10^{-9}$ | $2.9\times10^{-6}$ | $5.3\times10^{-3}$ | 11.0 |

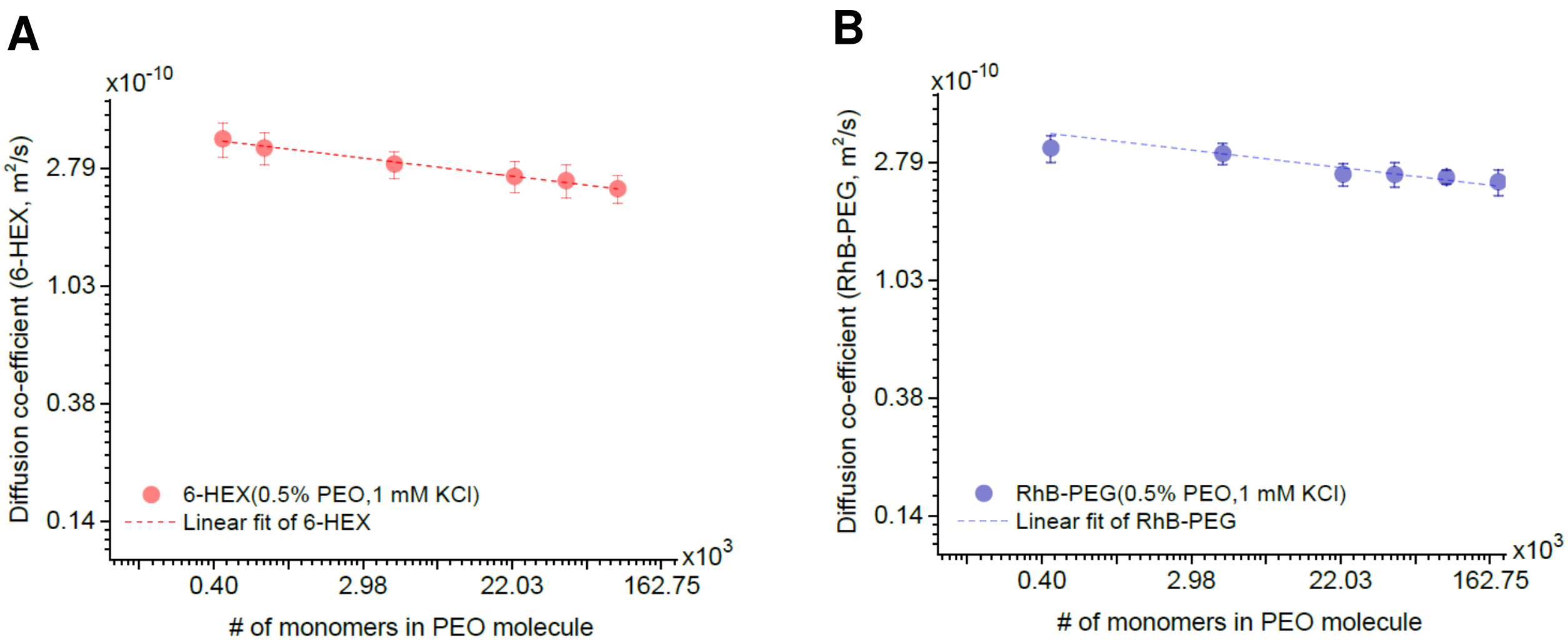

**Figure S4. Average diffusivity of non-interacting (A) 6-HEX dye molecules and (B) RhB - PEG (Rhodamine B tagged PEG of 1 kDa MW) in 0.5 wt% polyethylene oxide (PEO) solution at different MWs shows a linear pattern with a negative slope when plotted in log-log scale.** We plotted the diffusivity correlation $D_x = D_0\, e^{-N/N_c}$ with axes scaled to natural logarithm and calculated the resulting slope: $-\,1/N_c$ which solely depends on the properties of the polymer solution (PEO) and not on the diffusing species. All error bars represent standard deviations from at least 5 independent measurements.

Plotting natural logarithm of diffusion coefficients ($D_x$) of a non-interacting 6-HEX molecule with the natural logarithm of the number of monomers in the particular polymer of certain molecular weight, we can model the diffusivity through the following exponentially decaying correlation as described in eqn (4). Where, $D_0$ is the diffusion coefficient in pure water, $N$ is the number of monomers in the polymer chain of particular MW and $N_c$ is the characteristic number of monomers.

We found $N_c \sim 22$ from the experimental measurement of 6-HEX diffusion coefficients at 0.5 wt% PEO concentration. In other words, to attain an exponential decrease of diffusion coefficient, the number of monomers in the polymer should be $N = N_c$. Additionally, for non-interacting RhB-PEG polymer diffusion in 0.5 wt% PEO concentration, similar to the 6-HEX case, we found, $N_c \sim 22$ indicating $N_c$ might scale with polymer concentration and correlation length (mesh size) of the PEO polymer system.

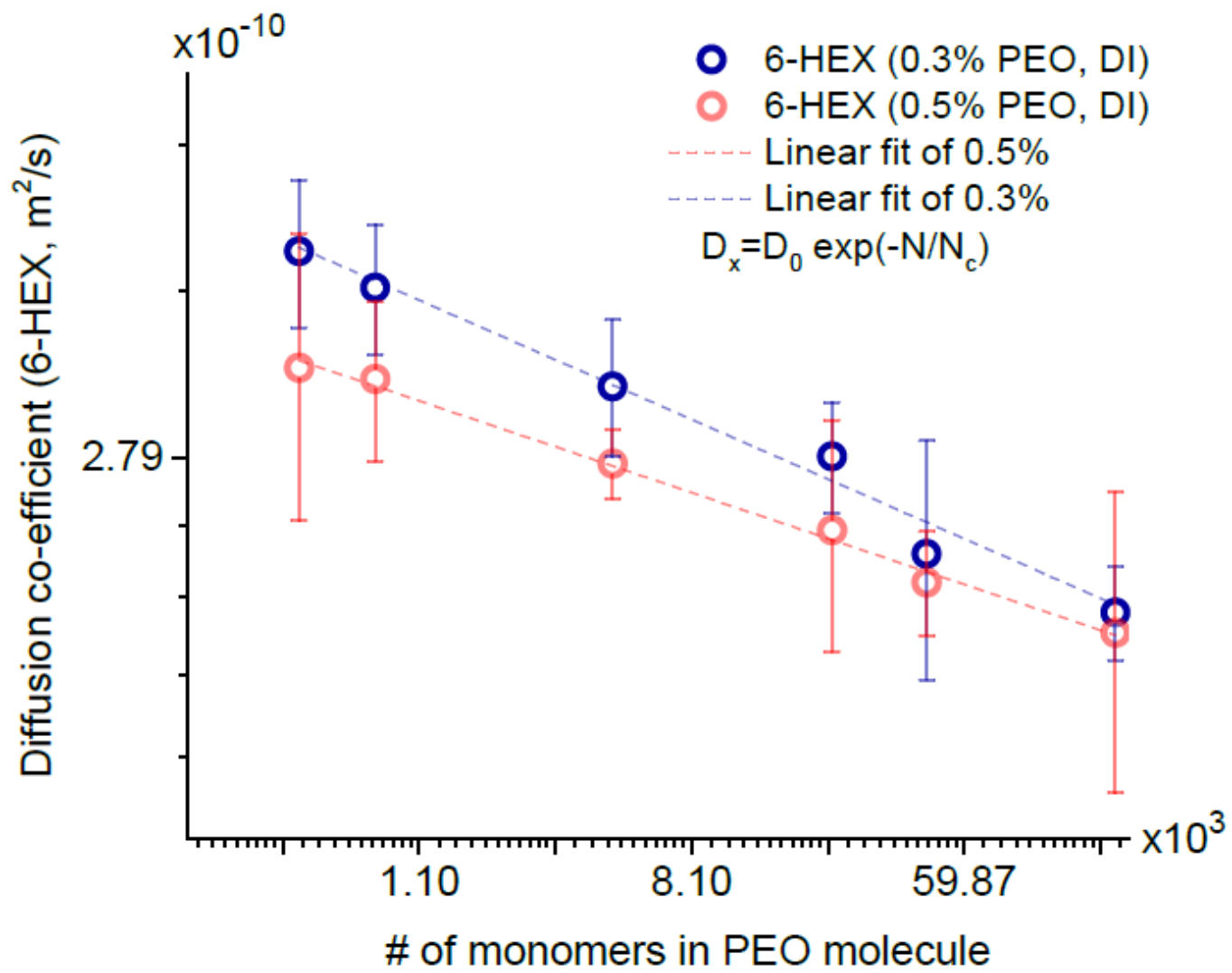

**Figure S5. Average diffusivity of non-interacting 6-HEX dye molecules in 0.3 wt% (blue circle) and 0.5 wt% (orange circle) PEO solutions at different N (# of monomers in 1 MW) shows different negative slopes in log-log scale.** We plotted diffusivity with y-intercept: $\ln D_0$ and slope: $-\frac{1}{k}\, C^{-\nu}$ in a log-log scale and calculated the average $k$ which mainly depends on the properties of the polymer solution (PEO). All error bars represent standard deviations from at least 5 independent measurements.

From experimental measurements of non-interacting 6-HEX diffusivities in 0.3% and 0.5% PEO solutions of different MWs, a scaling relation could be proposed between $N_c$ and the polymer concentration, $C$ such that-

$$N_c = k\, C^{\nu} \tag{5}$$

Where, k is a constant for the particular polymer system and $\nu$ is the Flory exponent. We can justify the above equation and most importantly, the scaling correlation $N_c \sim C^{\nu}$ through experimental data analysis using the table below (Supplementary Table S1).

We used a log-log linear fitting to estimate average values of $k$ and $N_c$ following the relation below which is derived from eqn (4) and eqn (5):

$$\ln D_x = \ln D_0 + \left(-\frac{1}{k}\, C^{-\nu}\right) N \tag{6}$$

From, experimental measurements, we found that average $k \sim 37.65$ for PEO (Supplementary Table S1).

**Table S2. Estimation and verification of proposed scaling: (i) $N_c = k\, C^{\nu}$ for 0.3% and 0.5% PEO solutions, (ii) $C^{\nu} \sim \xi^{1-3\nu}$ for 0.3% and 0.5% PEO solutions**

| MWs of PEO in Da | Correlation length at 0.3% ($\xi$) | Correlation length at 0.5% ($\xi$) | Ratio of correlation lengths raised to the power 1- 3$\nu$ ($\frac{\xi^{1-3\nu}\ at\ 0.5\%}{\xi^{1-3\nu}\ at\ 0.3\%}$) | Ratio of concentration raised to the power $\nu$ ($\frac{C^{\nu}\ at\ 0.5\%}{C^{\nu}\ at\ 0.3\%}$) | Ratio of $N_c$ ($\frac{N_c\ at\ 0.5\%}{N_c\ at\ 0.3\%}$) | Estimation of average k (with $C$ in wt%) |
|---|---|---|---|---|---|---|
| 2.00E+04 | 3.91E-08 | 2.64E-08 | 1.34E+00 | 1.34E+00 | 1.33E+00 | 37.65±0.3 |
| 3.50E+04 | 3.93E-08 | 2.65E-08 | 1.34E+00 | 1.34E+00 | 1.33E+00 | |
| 8.00E+04 | 3.97E-08 | 2.68E-08 | 1.34E+00 | 1.34E+00 | 1.33E+00 | |
| 1.00E+05 | 3.98E-08 | 2.68E-08 | 1.34E+00 | 1.34E+00 | 1.33E+00 | |
| 1.50E+05 | 3.99E-08 | 2.70E-08 | 1.34E+00 | 1.34E+00 | 1.33E+00 | |
| 2.00E+05 | 4.01E-08 | 2.70E-08 | 1.34E+00 | 1.34E+00 | 1.33E+00 | |
| 5.00E+05 | 4.04E-08 | 2.73E-08 | 1.34E+00 | 1.34E+00 | 1.33E+00 | |
| 8.00E+05 | 4.06E-08 | 2.74E-08 | 1.34E+00 | 1.34E+00 | 1.33E+00 | |
| 1.00E+06 | 4.07E-08 | 2.75E-08 | 1.34E+00 | 1.34E+00 | 1.33E+00 | |
| 1.50E+06 | 4.09E-08 | 2.76E-08 | 1.34E+00 | 1.34E+00 | 1.33E+00 | |
| 2.00E+06 | 4.10E-08 | 2.77E-08 | 1.34E+00 | 1.34E+00 | 1.33E+00 | |
| 3.00E+06 | 4.12E-08 | 2.78E-08 | 1.34E+00 | 1.34E+00 | 1.33E+00 | |
| 4.00E+06 | 4.13E-08 | 2.79E-08 | 1.34E+00 | 1.34E+00 | 1.33E+00 | |
| 6.00E+06 | 4.15E-08 | 2.80E-08 | 1.34E+00 | 1.34E+00 | 1.33E+00 | |
| 8.00E+06 | 4.16E-08 | 2.81E-08 | 1.34E+00 | 1.34E+00 | 1.33E+00 | |
| 1.00E+07 | 4.17E-08 | 2.82E-08 | 1.34E+00 | 1.34E+00 | 1.33E+00 | |

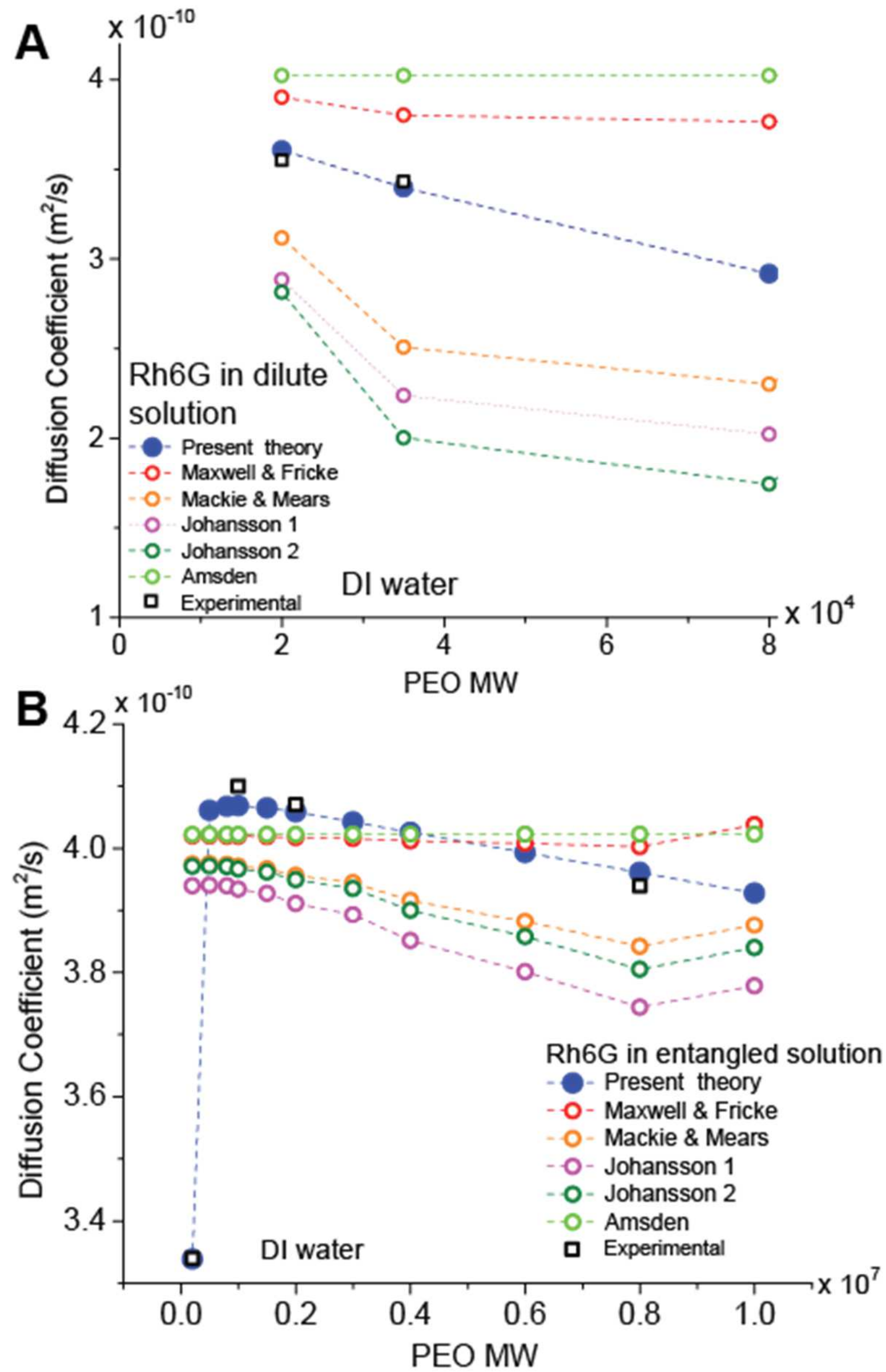

**Figure S6. Diffusivity of interacting Rhodamine 6G (Rh6G) dye molecules in 0.5 wt% polyethylene oxide (PEO) solution at different molecular weights shows varying patterns in (A) dilute and (B) semi-dilute entangled regimes, which cannot be explained by existing models and theories.** The proposed theory (filled blue circles) as presented in the manuscript, *w.r.t.* experimental data more accurately models the experimental observation of Rh6G diffusivity. The other models originating from hydrodynamic and/ or obstruction theory, such as Maxwell & Fricke [3], Mackie & Mears [4], Johansson model based on transport obstruction by polymer (Johansson 1[5]), Johansson model based on Brownian motion (Johansson 2[6]), and Amsden model based on Ogston's phenomenological approach[7], all deviate from experimental measurements of Rh6G diffusivity in both dilute and semi-dilute (entangled) solutions.

**Table S3. Different hydrodynamic models explaining diffusion of small molecules in polymer systems**

| Reference | System | Limitations |
|---|---|---|
| Cai and Rubinstein[8] | Nonsticky nanoparticles in polymer liquids | From small molecule perspective, this theory simply reinforces Stokes-Einstein diffusion coefficients and states that small molecules are not much affected by the polymers. |
| Yamamoto and Schweizer[9]<br>Yamamoto *et al*[10].<br>Carroll *et al.*[11] | Core-shell/vehicle model in polymer melts | Top-down approach of describing diffusion empirically using hydrodynamic and segmental contributions in polymer melts. |
| Holyst *et al*[12]. | Effective hydrodynamic radius dependent exponential diffusion model | Top-down diffusion model with multiple system/ probe specific parameters and follows an empirical approach. Probe size specificity, as the theory describes, is not warranted in the case of small particles/ molecules. |
| Kohli and Mukhopadhyay[13] | Different length scales for the diffusion of nanoparticles in semidilute polymer solutions | This approach is very similar to the empirical theory from Holyst *et al.* However, the length scale demarcation is quite wide (below and above radius of gyration) and the theory is particularly focused on the intermediate size range (larger than mesh size) of probes. The monotonic behavior of diffusivity was described with polymer volume fraction. |
| Senanayake *et al.*[14]<br>Nath *et al.*[15] | Hopping models | For small molecules, the hopping models are not effective as compared to larger sized probes. |

### S3. Hydrophobic force expression in terms of a pre-factor ($k_h$):

Hydrophobic force along x-direction ($F_h(x)$) is written according to Israelachvili and Pashley [16]:

$$F_h(x) = k_h e^{-x/x_0} \tag{7}$$

Where the pre-factor or adhesion force $k_h$ is related to the radius of the diffusing molecule ($a_m$) and a force constant $C$ (~ 0.1 N/m) such that $k_h = a_m C$. Additionally, $x$ represents the distance between two interacting surfaces and $x_0$ is the hydrophobic decay length characteristic of the system and spans around 1 nm [17]. We used $x_0$ as a parameter *w.r.t.* the degree of hydrophobicity in the system- 1.0 nm for PEO polymer systems in DI water[16]. In case of higher salt concentrations, $x_0$ is expected to decrease due to stronger hydrophobic interaction.

The hydrophobic interaction potential ($V_h(x)$) between two molecules can be expressed in the following integrated form:

$$V_h(x) = -\int_x^\infty F_h(x)\,dx = -\,k_h x_0 e^{-x/x_0} \tag{8}$$

Assuming that the sticking of a diffusing molecule to the polymer network is primarily due to hydrophobic interaction, the sticking or weak binding free energy ($\Delta G^0 = -RT\ ln(c^0/K_d)$) should closely follow hydrophobic potential at $x_0 = 0$. Therefore $k_h$ could be expressed as:

$$k_h = \frac{1}{x_0}\ k_B T\ ln\left(\frac{C^0}{K_d}\right) = \frac{1}{x_0} K_B T\ ln(c^0 K_a)$$

Where $k_B T$ is the thermal energy with Boltzmann constant $k_B$, $c^0$ = 1 M, $K_a$ is the equilibrium sticking constant, $K_d$ is the equilibrium escape constant and absolute temperature is T.

**Table S4. Interaction parameters of Rh6G and Ficoll in DI water across all PEO conc. regimes**

| Diffusing molecule | $\gamma_s$, N-s/m | $k_h$, N | $x_0$, m | $\tau_C \sim \frac{\gamma_s x_0}{k_h}$, s |
|---|---|---|---|---|
| Rh6G | $3.0\times10^{-5}$ | $3.2\times10^{-11}$ | $1.0\times10^{-9}$ | $9.6\times10^{-4}$ |
| Ficoll | - | - | - | $8.6\times10^{-4}$ |

Note: 1. $k_h$ of Rh6G was evaluated by measuring equilibrium association constant ($K_a$)[1]
2. $\gamma_s$ was theoretically evaluated using Vogel-Fulcher equation[18] (see SI, section S7.1)

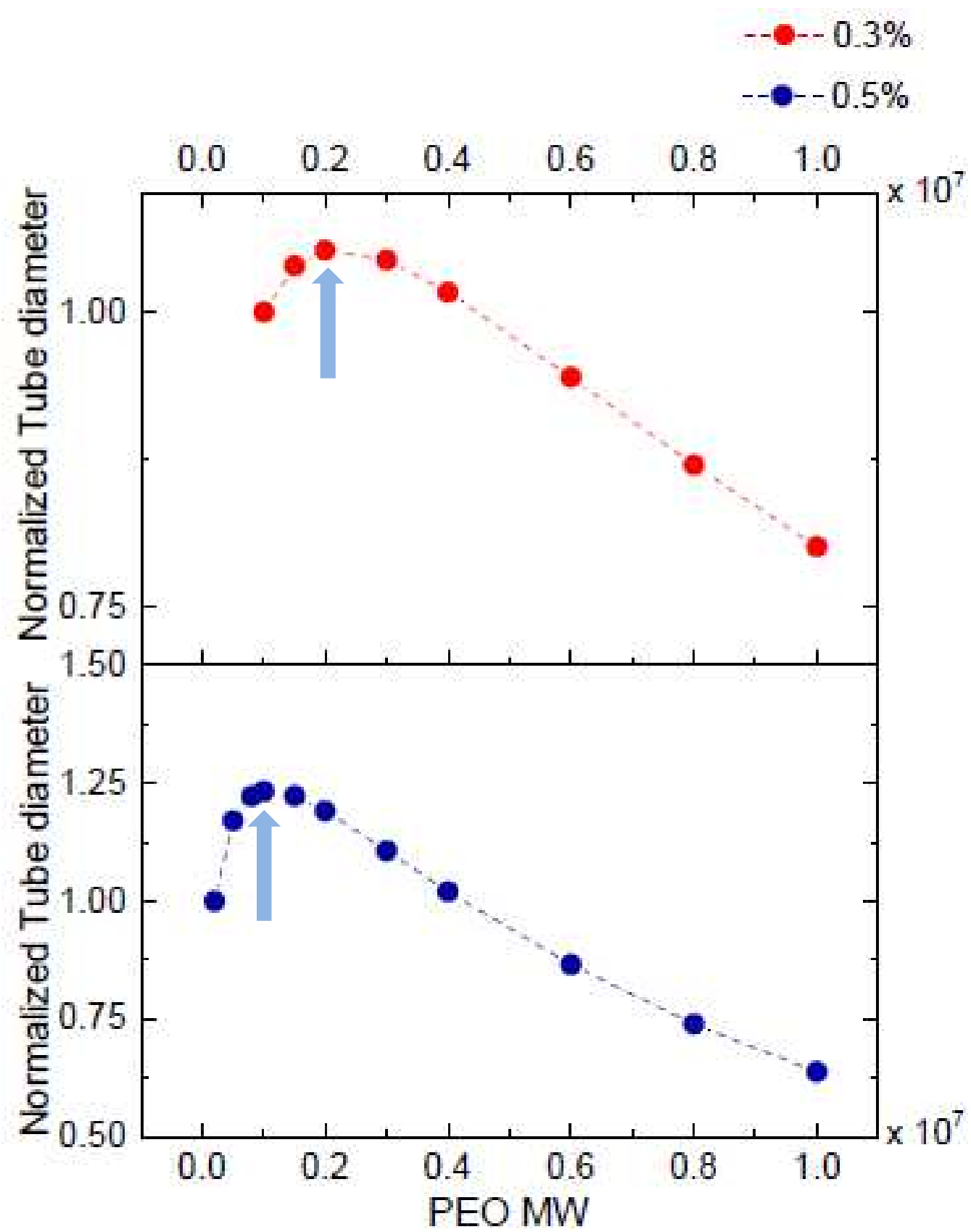

**Figure S7. Theoretical estimation of normalized tube diameter of PEO solutions at 0.3% and 0.5% concentrations show that tube sizes are at maxima at 2000K and 1000K MWs of PEO, respectively.** This observation supports the hypothesis that diffusion is facilitated within tube compartments in the semi-dilute and entangled polymer solution. Additionally, this observation also explains the local maxima of the diffusion coefficient of interacting molecules (like Rh6G) in PEO solutions at respective concentrations. The normalizations were performed *w.r.t.* tube diameters at 1000K and 200K PEO MWs at 0.3% and 0.5%, respectively. The arrows (blue) indicate the positions of maxima of normalized tube diameter at respective concentrations.

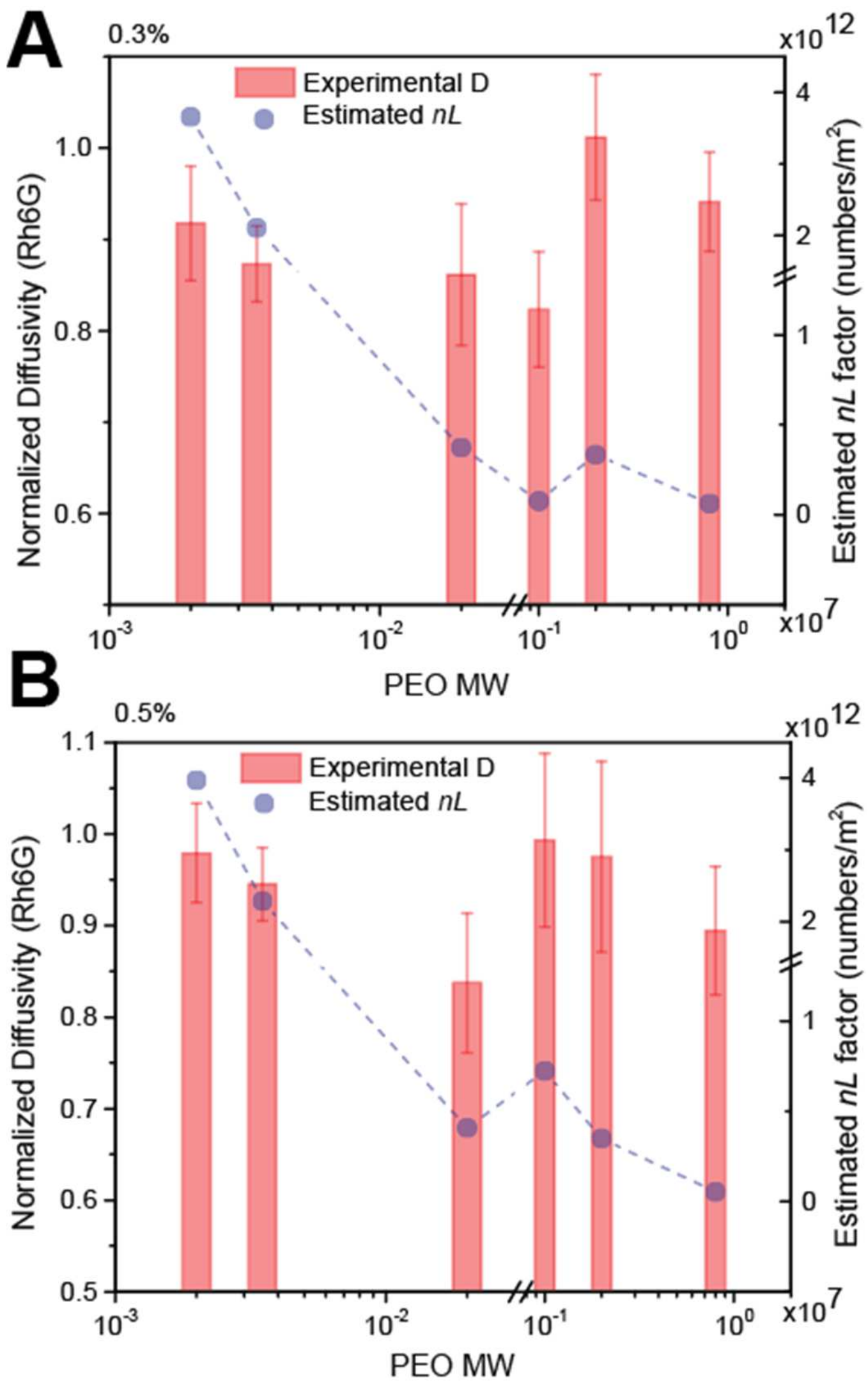

**Figure S8. Theoretical estimation of surface density (number/m$^2$) of polymer chains or effective crowding density ( the *nL* factor) of PEO solutions at (A) 0.3% and (B) 0.5% concentrations in DI water shows that local optimum of diffusivity depends on *nL*.** This observation supports the hypothesis that mobility is facilitated by the arrangement of polymer chains itself and in the dense entangled regime, it is still possible to enhance mobility through adjustments in polymer concentrations and length scales. Under the conditions of the experiments, we found that the measured local maximum of the diffusion coefficient in the entangled regime corresponds to the local maximum of *nL.*

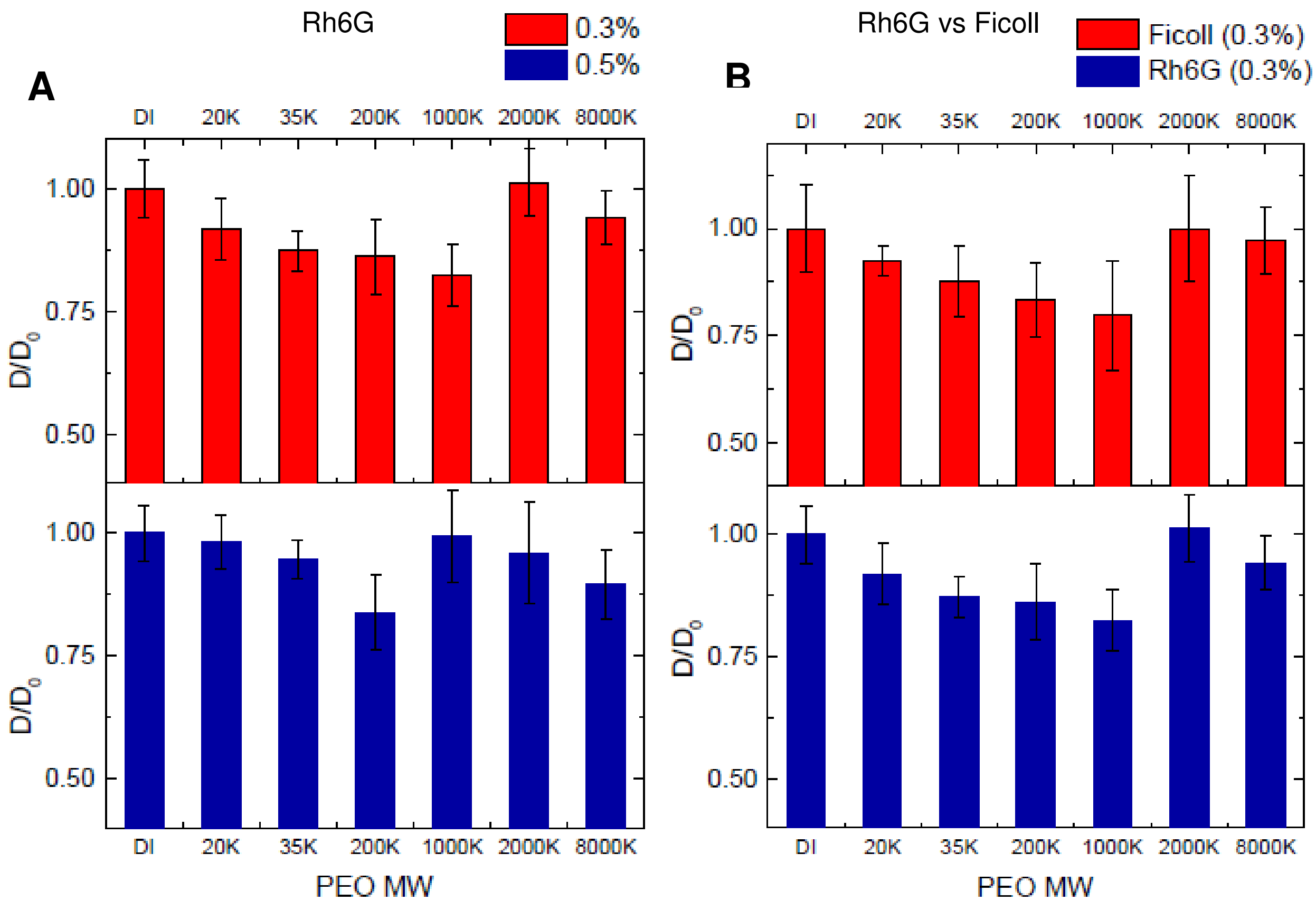

**Figure S9. Variation of normalized diffusion coefficient ($D/D_0$) of Rh6G molecule diffusing in 0.3 wt% (red) and 0.5 wt% (blue) PEO solutions of different MWs and comparison with Ficoll diffusing in 0.3 wt% (red) PEO solutions.** (A) The initial decrease of diffusion coefficient continues up to 1000K PEO in the case of 0.3% solution and a similar decrease continues up to 2000K PEO in the case of 0.5% solution. The maxima of the diffusion coefficient in both cases corresponds to maxima in the tube size of the polymer network (Figure S7). (B) A similar trend of diffusion coefficient exists for interacting Ficoll molecules at different MWs of PEO solutions, as compared to interacting Rh6G molecules under similar conditions. All error bars are standard deviations from at least 5 independent measurements.

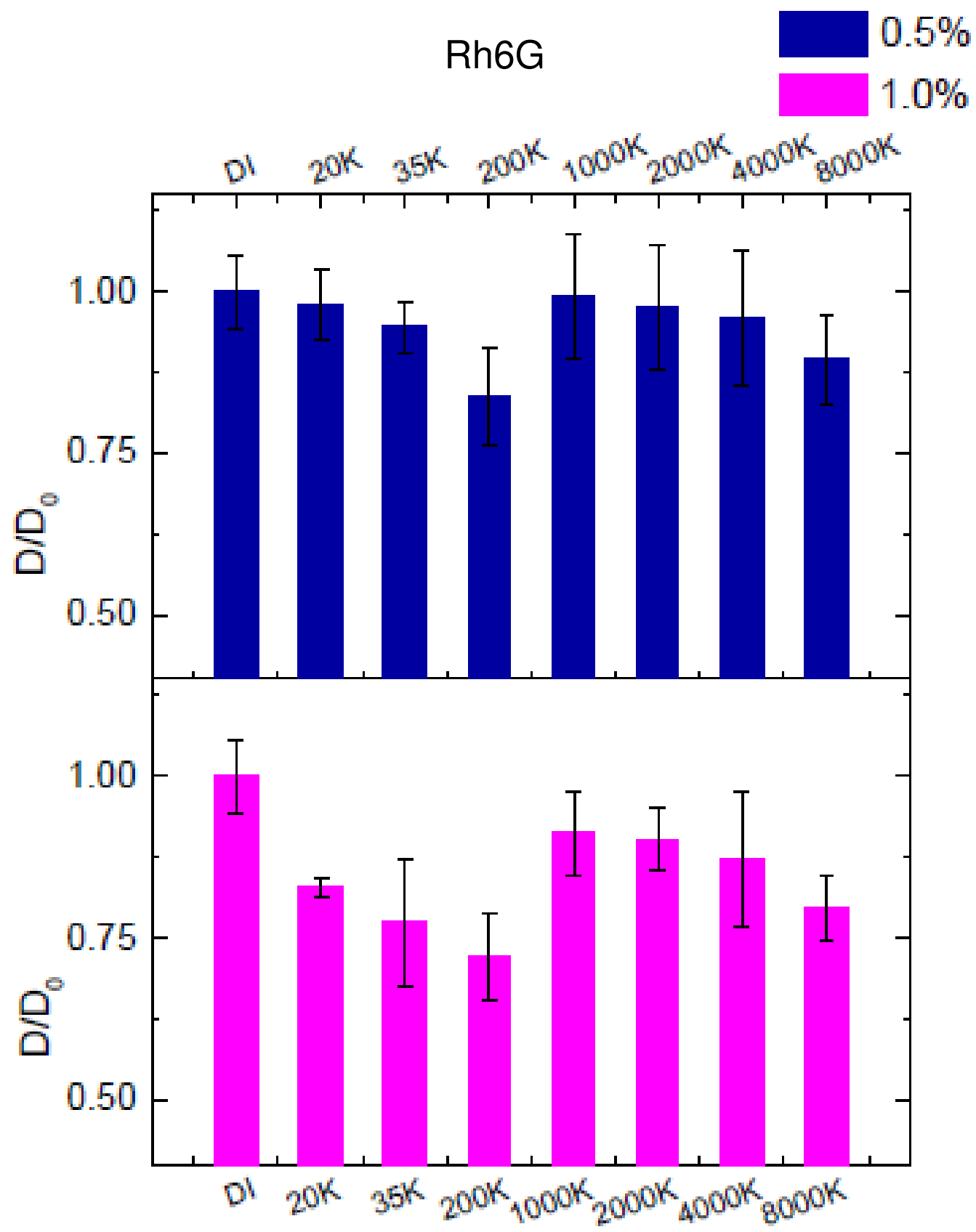

**Figure S10. Variation of normalized diffusion coefficient ($D/D_0$) of Rh6G molecule diffusing in 0.5 wt% (blue) and 1.0 wt% (magenta) PEO solutions of different MWs shows that decrease of diffusion coefficient continues till 200K PEO in both cases of PEO concentrations.** However, due to higher polymer concentration and therefore, higher frictional effects in 1% PEO solution, the measured diffusivities are smaller compared to the 0.5% PEO case. All error bars are standard deviations from at least 5 independent measurements.

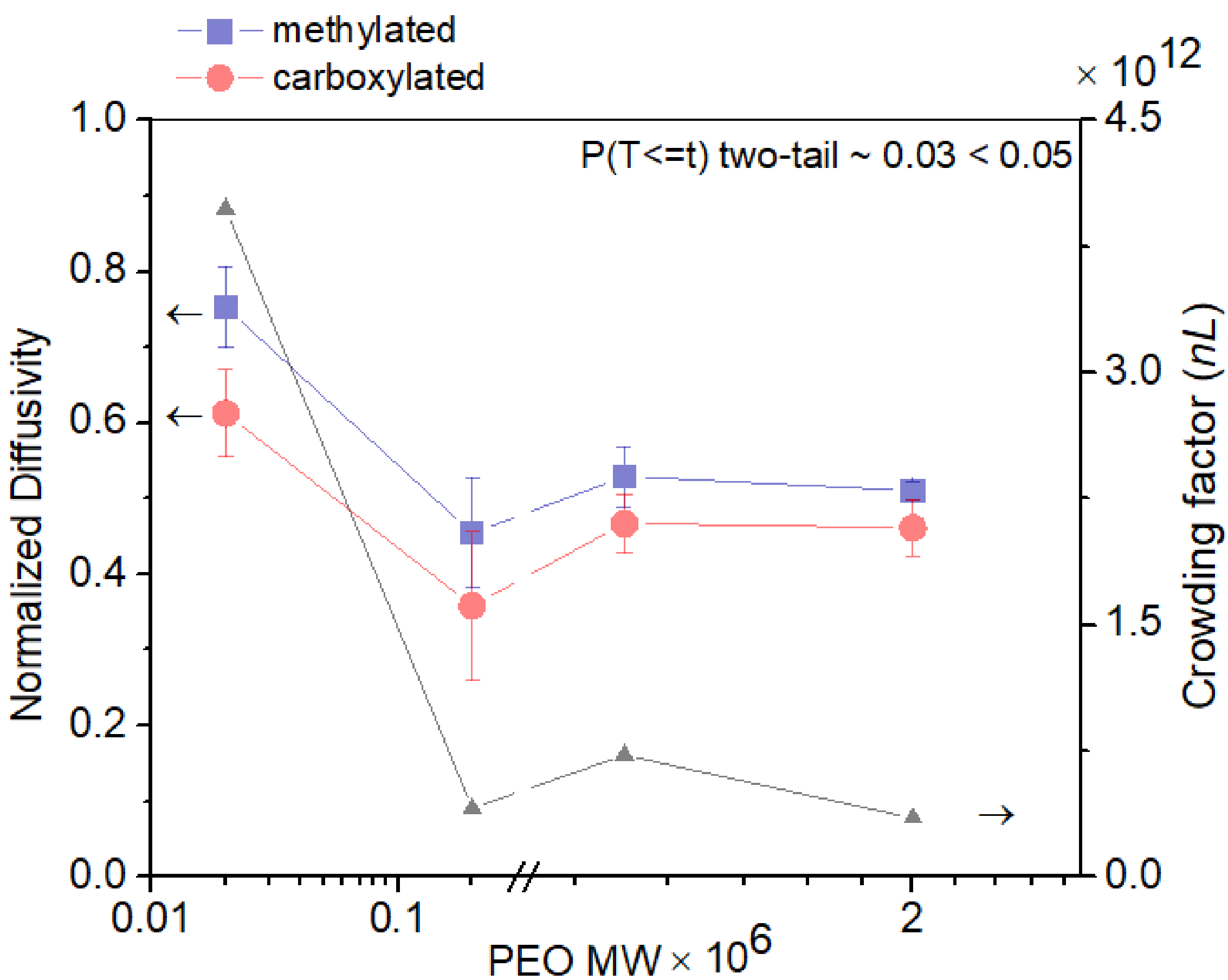

**Figure S11. Measurement of diffusion coefficients of 5 nm AuNPs coated with 5 kDa long chains of methylated PEG (blue squares) and carboxylated PEG (red circles) in 0.5 wt% PEO solutions at different MWs shows that normalized diffusion coefficients are higher and statistically significant (P(T<=t) ~ 0.03 < 0.05 following paired 2-sample t-test on mean) in the former case. The normalized diffusivity values correlate to the crowding factor, *nL*, in both cases.** We used methylated (CGM5K) and carboxylated (CGC5K) AuNPs dispersed in DI water (PDI < 0.1) from NNCrystal Corporation, Fayetteville, AR. We used a dynamic light scattering (DLS) method using λ ~ 633 nm laser light source and $\theta \sim 90°$ scattering angle. The intensity correlation functions obtained from DLS were fitted using a method described by Carroll *et al.*[11] at small time intervals (~ 400 μs) to estimate diffusion coefficients of AuNPs. All error bars are standard deviations from at least 4 independent measurements.

**Table S5. MSD Comparison of proposed compartment models at small times ($\tau_s \ll t \ll \tau_R$)**

| **Model_1 : Langevin** | **Model_2: Fokker-Planck** |
|---|---|
| Complete MSD model: (similar to core-shell/ vehicle model)[11]<br><br>$MSD\|_{total} = \{MSD\|_{unbound}\} + \{MSD\|_{bound}\}$<br><br>$= 6\,[\{\,D\,t + \frac{1}{2}\left(\frac{x(0)}{\tau_c}\,t\right)^2\} + \left\{\frac{1}{2}\,b^2 N_e^{\frac{1}{2}} \tau_s^{-\frac{1}{4}} t^{\frac{1}{4}}\right\}]$<br><br>Where, $D$ is the diffusion coefficient of the molecule based on the microviscosity of the solution, $b$ is the segmental length of the polymer strand, $N_e$ is the number of monomer units in the entanglement length and $\tau_s$ is the segment relaxation time. The above model applies when time ($t$) is much less than Rouse time ($\tau_R$), i.e. $t \ll \tau_R$. | Complete MSD model: (similar to core-shell/ vehicle model) [11]<br><br>$MSD\|_{total} = \{MSD\|_{unbound}\} + \{MSD\|_{bound}\}$<br><br>$= \{3\int_{-\infty}^{\infty} r^2 P(r,t)\,dr\} + 6\left\{\frac{1}{2}\,b^2 N_e^{\frac{1}{2}} \tau_s^{-\frac{1}{4}} t^{\frac{1}{4}}\right\}$<br><br>$= \{3\,x_0^3 \int_{-\eta_t}^{\eta_t} \eta^2 P(\eta,\tau)\,d\eta\} + 6\left\{\frac{1}{2}\,b^2 N_e^{\frac{1}{2}} \tau_s^{-\frac{1}{4}} \tau_c^{\frac{1}{4}}\,\tau^{\frac{1}{4}}\right\}$<br><br>$= 6\,[\{\,x_0^3 \int_0^{\eta_t} \eta^2 P(\eta,\tau)\,d\eta\} + \left\{\frac{1}{2}\,b^2 N_e^{\frac{1}{2}} \tau_s^{-\frac{1}{4}} \tau_c^{\frac{1}{4}}\,\tau^{\frac{1}{4}}\right\}]$<br><br>Where, $P(\eta,\tau)$ is the dimensionless probability distribution function and $\eta_t = (a_t/2)/x_0$ is the dimensionless compartment or cage boundary. |
| Simplified MSD model as used in this work:<br><br>As the bound contribution from polymer dynamics subject to reptation is weaker, we can simplify the above MSD as follows: $MSD \approx 6\,[\,D\,t + \frac{1}{2}\left(\frac{x(0)}{\tau_c}\,t\right)^2]$ at small times | Simplified MSD model as used in this work:<br><br>$MSD \approx 6\,x_0^3 \int_0^{\eta_t} \eta^2 P(\eta,\tau)\,d\eta$ |

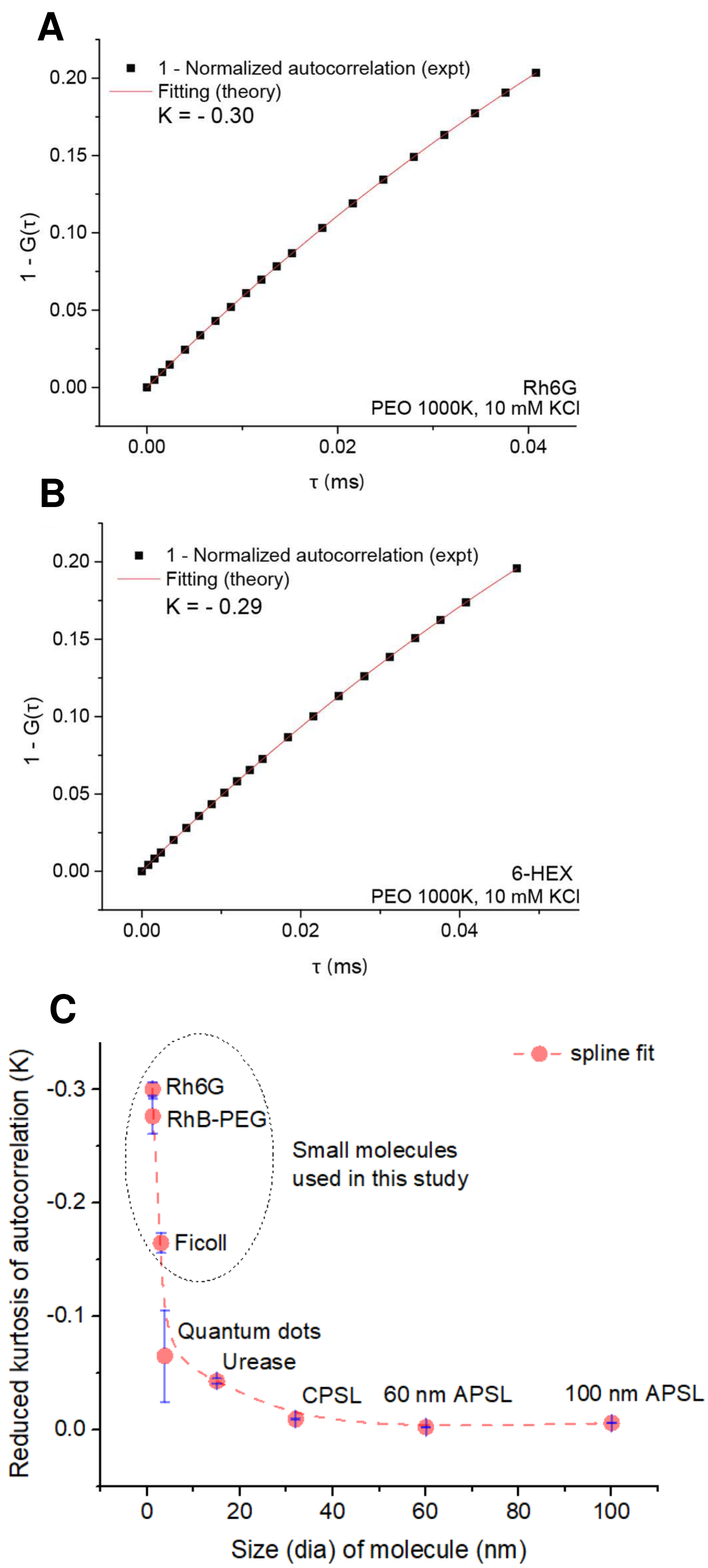

**Figure S12. Fitting of 1-$G(\tau)$ vs $\tau$ in cases of (A) Rh6G diffusion and (B) 6-HEX diffusion in 0.5 wt% PEO solution of MW 1000K with 10 mM KCl to estimate the reduced kurtosis (K) of the propagator (probability distribution)** [2]**.** Fitting with 1-$G(\tau) = -c_2 t^{\alpha} - c_4 t^{2\alpha}$ within short span of time (when $G(\tau)$ decays to 0.8), we estimated the reduced kurtosis for both Rh6G and 6-HEX molecules are $K \sim -\ 0.3$ which corresponds to deviation from Gaussian nature of their propagators and therefore, displacements. (C)

Dependence of reduced kurtosis of autocorrelation with molecular sizes of different diffusing species in the solution of 0.3 wt% PEO of MW 1000K at a small time interval of ~ 0.05 ms. The Gaussian nature is attained following a decaying hyperbolic distribution pattern with molecular size. The deviation from Gaussianity is mostly observed around the size range of ~3 nm for particles subject to similar conditions. CPSL: negatively charged particles and APSL: positively charged particles. All error bars are standard deviations from 3 independent measurements.

**S4. Derivation of boundary condition (Model_2) on the polymer strand at equilibrium:**

Within the control volume of the compartment $V_t$, we assume that $N_D$ is the number of monomer sites occupied by dye molecules of total number concentration $C_D$ such that the probability of occupied sites, $P_D = \frac{N_D/V_t}{C_{monomer}}$, where $C_{monomer}$ is the total number of sites on the polymer segment available to dye molecules to stick (Schematic S2). Therefore, the number concentration of free dye molecules, $C_D|_{free} = C_D - \frac{N_D}{V_t}$, where $N_D/V_t$ represents the number concentration of bound dyes. If we further assume that the dye sticking process is a first order phenomenon with rate constant $k_1$ and dye escape process is also a first order phenomenon with rate constant $k_2$, we can formulate the below equation keeping a track of site occupancy through a balance between sticking and escaping phenomena (Schematic S2):

$$\frac{dN_D}{dt} = k_1 \, C_D|_{free} \, V_t - k_2 \, C_D|_{occipied} \, V_t \quad (9)$$

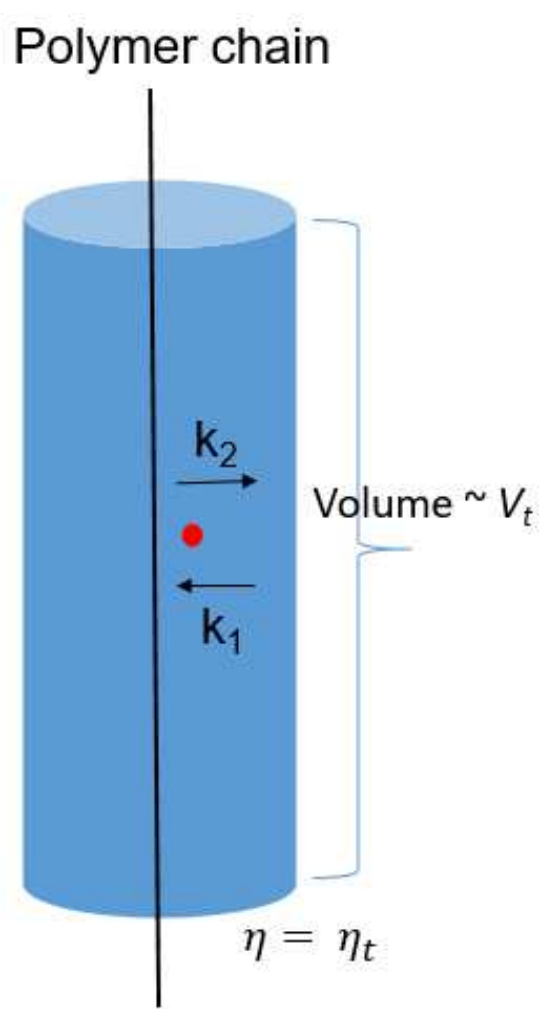

**Schematic S2. Kinetic representation of polymer site occupancy by sticky molecules and its corresponding probability density on a polymer chain.**

Where $C_D|_{occipied} = N_D/V_t$. The characteristic sticking time, $\tau_B \sim 1/k_1$ and the desorption time, $\tau_{uB} \sim 1/k_2$. In the above equation, we can replace $C_D|_{free}$, $k_1$ and $k_2$ along with the application of $P_D = \frac{N_D/V_t}{C_{monomer}}$ such that the differential equation takes the following form:

$$\frac{dP_D}{dt} + \left(\frac{1}{\tau_B} + \frac{1}{\tau_{uB}}\right) P_D = \frac{P_0}{\tau_B} \quad (10)$$

Where, $P_0 = C_D/C_{monomer}$. In the above equation, subject to steady-state and when $\tau_{uB} \gg \tau_B$ (sticky interaction dominates in the time interval of measurements), we can simplify $P_D \approx P_0$.

Along the one dimensional x-axis (or radial r-axis), assuming symmetric conditions *w.r.t.* the polymer strand, the probability density function ($P(x,t)$) could be written as: $\int_{-a_m}^{a_m} P(x,t)\, dx = 2\, P_0$. In dimensionless terms, $P(\eta,\tau)|_{\eta \to \eta_m} = P_0/a_m$ is the boundary condition on the polymer strand at the closest approach of the molecule whose radius is $a_m$. This is also the result of the assumption that probability density across the small thickness of the polymer strand stays constant. Please note the dimensionless parameters in the boundary condition are dimensionless time $\tau = t/\tau_C$ and dimensionless position $\eta = x/x_0$.

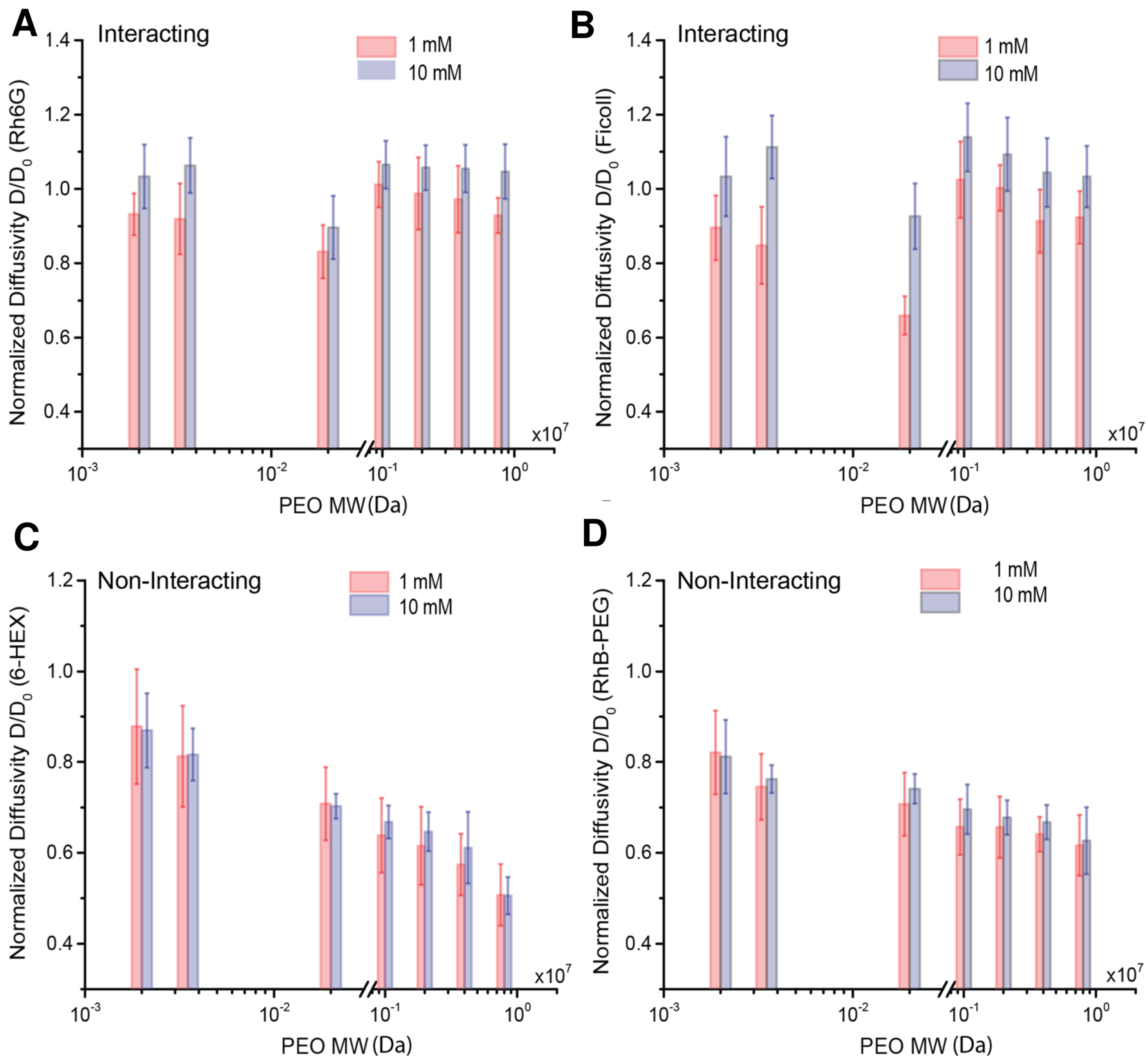

**Figure S13. The normalized diffusion coefficient pattern of sticky molecules (Rh6G, Ficoll) is different from inert molecules (6-HEX, RhB-PEG) such that at higher salt concentrations (10 mM) diffusion coefficient of sticky molecules is enhanced with a concomitant change in mobility pattern with increasing MWs (Da) of polymer.** (a) Comparison of normalized Rh6G diffusion coefficients in 0.5% PEO solution with 1 mM (red bars) and 10 mM KCl (blue bars) concentrations at different MWs of PEO. (b) Comparison of normalized Ficoll diffusion coefficients in 0.5% PEO solution with 1 mM (red bars) and 10 mM KCl (blue bars) concentrations show patterns similar to Rh6G. (c) Comparison of normalized 6-HEX diffusion coefficients in 0.5% PEO solution with 1 mM (red bars) and 10 mM KCl (blue bars) concentrations show similar patterns irrespective of added salt concentrations. (d) Comparison of normalized RhB-PEG diffusion coefficients in 0.5% PEO solution with 1 mM (red bars) and 10 mM KCl (blue bars) concentrations show patterns similar to 6-HEX. All error bars are standard deviations from at least 5 independent measurements.

**Table S6. Estimation of crowding factor $nL$ for different MT systems**

| # | Systems | Estimated $nL$ ($\times 10^{14}$ #/$m^2$) | Calculation details |
|---|---|---|---|
| 1 | Organism: *Pisaster ochraceus*<br><br>Evolution years in millions of years ago (MYA): Evolution of Asteroidea (like *Pisaster ochraceus*) took place ~ 500 MYA when *P. ochraceus* diverged from other sea stars [19,20]. | 3.7 | Meiosis I metaphase spindle of sea star *P. ochraceus* oocyte was studied by Sato *et al.* using fiber induced birefringence. The reported microtubule density was 106 $\#/\mu m^2$. The reported spindle width (*W*) was ~8 µm and pole to pole spindle length ($L_{sp}$) was ~10 µm. We would use the ellipsoidal model of the spindle to calculate area (*A*) and volume (*V*), respectively [21].<br><br>Fiber density ($\#/m^3$) $n = 106A/V$, where $A \sim 235.3\ \mu m^2$ and $V \sim 335.1\ \mu m^3$ as per the ellipsoidal structure of the spindle. Therefore, $nL = 3.7 \times 10^{14}$ $\#/m^2$. Please note that the MT fiber length (*L*), is around half of the spindle length, i.e. $L \sim L_{sp}/2$. |
| 2 | Organism: *Arbacia lixula*<br><br>Evolution years (MYA): Evolution of Echinoidea (like *Arbacia lixula*) took place ~ 450 MYA and diverged from closely related species [19,22].<br><br>Note that the "9+2" axoneme structure is much more primordial and has been in use since last 1 billion years from diversification of LECA [23]. | 3.3 | Sea urchin *A. lixula* sperm flagella is arranged according to the classic "9+2" axoneme pattern with a central pair and nine outer doublets. These structures are evolutionary conserved and started to exist 1 billion years from now, since the diversification of last eukaryotic common ancestor (LECA) [23,24].<br><br>Using cryo-electron microscopy, Linck *et al.* collected high resolution images of "9+2" organization. 20 MTs were arranged in the cross-section. The diameter of the cylindrical structure was ~280 nm and the MT fibers which spanned the sperm flagella length (*L*) was ~35 µm long. Therefore, $nL = \frac{20}{\pi\,(0.14\times10^{-6})^2\,35} 35 = 3.25 \times 10^{14}\ \#/m^2$ [25].<br><br>Another related structure was "9+0" axoneme where the central MT pair is missing and the circumferential MTs are arranged into 9 triplets. Such structures more often remain non-motile and used for sensory purposes. We estimate, $nL = \frac{27}{\pi\,(0.14\times10^{-6})^2\,L} L = 4.4 \times 10^{14}\ \#/m^2$ for "9+0" architecture, assuming the same geometric organization as before [26]. |
| 3 | *Organism: Saccharomyces cerevisiae*<br><br>Evolution years (MYA): ~ 330 MYA when *S. cerevisiae* diverged from *S. pombe* [27]. | 4.2 | The spindle parameters of buddying yeast, *S. cerevisiae* were reported as follows: spindle length $L_{sp} \sim 2$ µm, the area of spindle pole body (SPB): $A \sim 20106\ \text{nm}^2$. SPB organizes a total of ~36 kinetochore and interpolar MTs [28,29]. |
| | | | As per Inuoue and Sato [30], spindle volume ($V$) with SPB calculated as $V \sim \frac{4}{3}LA$. Where $L$ is half spindle length and $A$ is area of SPB. Therefore, $nL = \frac{36}{V} = 4.2 \times 10^{14}$ $\#/m^2$. |
| 4 | *Organism: Caenorhabditis elegans*<br><br>Evolution years (MYA): ~ 100 MYA when *C. elegans* diverged from *C. briggsae* [31–34]. | 3.4 | Using electron tomography on *C. elegans* embryo, Redemann *et al.* found spindle width (W) ~6 µm and half spindle length ($L$)~6 µm [35]. Additionally, we estimated the length average, area based number density of kinetochore and spindle MTs which formed the main metaphase spindle, ~65#/ µm$^2$ [35].<br><br>Therefore, fiber density ($\#/m^3$), $n = 65A/V$, where $A \sim 193.13\, \mu m^2$ and $V \sim 226.19\, \mu m^3$ as per the ellipsoidal structure of the spindle. Therefore, $nL$= $3.4 \times 10^{14}\ \#/m^2$. |
| 5 | *Organism: Drosophila melanogaster*<br><br>Evolution years (MYA): ~ 30.5 MYA, when *D. melanogaster* diverged from *scaptomyza* species [36]. Whereas divergence time between *melanogaster* and *obscura* group was estimated ~ *30-35 MYA* [37]. Similar timeline (~ 25-30 MYA) was reported by Levine *et. al.* based on a different marker [38]. | 4.3 | The number density of MT filaments in metaphase spindle of *Drosophila* S2 cells is difficult to ascertain experimentally due to bundle formation with varying numbers of MT filaments [39]. However, when the tubulin dimer concentration, $C_d$ (mg/ml) is known, one can estimate the number concentration $n$ using a simple formula presented by Brown and Berlin [40] for MT system of average filament length of $L$ µm as follows:<br><br>$n = \frac{C_d}{110 \times 10^3} \times \frac{N_A}{1625\, L} \approx \frac{3.37 \times 10^{18}}{L} C_d\ \#/m^3$<br><br>The above formulation assumes 1625 dimers per µm of MT length and 110 kDa is the molecular weight of tubulin dimers. Additionally, if $\rho_L$ is the mass of tubulin monomer per unit length of the monomer, then the mesh size $\xi_m$ in a cubic lattice of MTs is given in terms of tubulin monomer concentration $C_m$ as below [41]:<br><br>$\xi_m = \sqrt{\frac{3\, \rho_L}{C_m}} \approx \frac{0.8}{\sqrt{C_m}}$ (in µm)<br><br>Using ~ 50 nm as the average distance between MTs within bundles of *Drosophila* metaphase spindles [39], i.e. $\xi_m \sim 50\, nm$, we could estimate $C_d$ ~128 mg/ml.<br><br>Therefore, for an average MT length of $L$ µm, we could easily estimate $nL \sim \frac{3.37 \times 10^{18}}{L} 128\, L \times 10^{-6}\ \#/m^2 \sim 4.3 \times 10^{14}\ \#/m^2$ , which is near to the average $nL$ value in MT system ($nL \sim 3.7 \times 10^{14}\ \#/m^2$). |
| 6 | *Organism: Xenopus laevis*<br><br>Evolution years (MYA): ~ 18 MYA as per allotetraploid event and it diverged from *X. borealis* [42]. | 4.4 | Kaye *et al.* used a combination of fluorescent lifetime imaging-Forster resonance energy transfer based method to evaluate the tubulin concentration profile in the metaphase spindle of *X. laevis* egg extract [43].<br><br>We know from electron micrograph analysis of helical MT fiber [44] that ~ 30 tubulin monomers translate into 14 nm length. Therefore, within a MT seed of size 50 nm [43], the number of tubulins, $n_{tub} = \frac{50\times10^{-9}}{14\times10^{-9}} \times 30$. With an estimated mean of tubulin concentration, $C{\sim}16$ µM [43], within MTs of the inner spindle, the number concentration ($\#/m^3$) of MT fibers $n = CN_A/n_{tub} = 8.9 \times 10^{19}\ \#/m^3$. Therefore, with typical MT fiber length ($L$) of 5 µm, $nL = 4.4 \times 10^{14}\ \#/m^2$. |
| 7 | *Organism: Homo sapiens*<br><br>*Cell:* HeLa cell<br><br>Evolution years (MYA): Assuming divergence and appearance of recent rare alleles (including *BRCA1* and *BRCA2* loss-of-function mutations) starting from ~ 10K years ago with the advent of agriculture and settlement [45]. | 5.4 | Nixon *et al.* reported the MT fiber density in mitotic spindle of HeLa cells ~210 $\#/\mu m^2$[46]. Jordan *et al.* reported the spindle width (W) as ~9.6 µm and interpolar distance ($L_{sp}$) was ~7.4 µm [47].<br><br>Therefore, fiber density ($\#/m^3$), $\nu = 210A/V$, where $A{\sim}\ 246.5\ \mu m^2$ and $V{\sim}357.1\ \mu m^3$ as per the ellipsoidal structure of the spindle. Therefore, the estimated $nL = 5.4 \times 10^{14}\ \#/m^2$.<br><br>Interestingly, with the addition of Taxol, the spindle assembly was stabilized and no change in MT mass was observed until 10 nM. For example, after 3 nM and 10 nM Taxol additions, the interpolar distances were reduced to ~6.1 µm and ~4.0 µm, respectively [47]. Concomitantly, the $nL$ parameter was reduced from $4.8 \times 10^{14}\ \#/m^2$ to $4.0 \times 10^{14}\ \#/m^2$, respectively. The detail values are reported in Supplementary Table S4 below. |
| 8 | *Metaphase spindle with SAC silencing (spindle assembly checkpoint)* | 3.0 | Modeling of spindle dimensions *w.r.t.* robust SAC silencing was performed by Chen and Liu [48]. Their model of spindle geometry was conical.<br><br>Using spindle width ($W$) ~9.6 µm and spindle length ($L_{sp}$) ~12 µm of a normal human cell [49], we found the aspect ratio $W/L{\sim}0.8$ which corresponds to active SAC silencing and mitotic progression.<br><br>The MT density from their modeling result was ~55 $\#/\mu m^2$ at ~9.6 µm spindle width.<br><br>Using spindle dimensions, the slant height of the cone, $l = \sqrt{(L)^2 + (W/2)^2} = 7.7\ \mu m$. Note that the MT fiber length- $L{\sim}L_{sp}/2$. |

| | |
|---|---|
| | Furthermore, the surface area ($A = 2\pi l\, W/2 \approx 233\ \mu m^2$) and volume ($V = \frac{2}{3}\pi\left(\frac{W}{2}\right)^2\left(\frac{L}{2}\right) \approx 289\ \mu m^3$) of the conical spindle gave the MT number concentration in the spindle as $n = 55A/V \approx 44.4\ \#/\mu m^3$. Therefore, $nL \approx 3.0 \times 10^{14}\ \#/m^2$. |

**Table S7. $nL$ estimation of HeLa cell metaphase spindle with and w/o taxol treatment** [47]

| Taxol conc. (nM) | $nL$ ($\times 10^{14}\ \#/m^2$) | $r_{nL} = \frac{nL\ of\ the\ system}{nL\ for\ mobility\ maximization}$ |
|---|---|---|
| 0 | 5.4 | 1.6 |
| 3 | 4.8 | 1.4 |
| 10 | 4.0 | 1.1 |

## S5. FCS data analysis:

### S5.1. Time span of diffusion analysis and MSD inversion from FCS data:

According to Stoelle and Fradin[50], as long as the particle diffusion time ($\tau$) remains much smaller than the escape time from the observation volume ($\tau_{1/2}$), i.e. $\tau \ll \tau_{1/2}$, actual MSD is returned regardless of the Gaussianity and MSD can be inverted in its relation to the autocorrelation $G(\tau)$, even if the propagator is not Gaussian. For example, in case of a particle diffusing at $D \sim 10^{-10}\ m^2/s$ within a confocal beam waist radius of $\omega \sim 600\ nm$, the time scale of $\tau_{1/2} \sim \omega^2/D \approx 4\ ms$. Therefore, we used our analysis time, $\tau \sim 0.4\ ms$ which satisfies the criteria of $\tau \ll \tau_{1/2}$.

### S5.2. Fluorescence correlation spectroscopy (FCS) setup and analysis:

For carrying out FCS measurement a custom-built optical setup was used which consisted of an excitation source from a PicoTRAIN 532 nm, 80 MHz, 5.4 ps pulsed laser (High-Q Laser). Samples were excited by focusing the laser through Olympus 60× 1.2-NA water-immersion objective which is attached to IX-71 microscope (Olympus). A dichroic beam splitter (Z520RDC-SP-POL, Chroma Technology) was used to separate the emission signal from the excitation beam. The former was focused through a confocal pinhole consisting of a 50 μm, 0.22-NA optical fiber from Thorlabs. The signal detected by a photomultiplier tube was routed to a preamplifier (HFAC-26) and coupled to a time-correlated single-photon counting (TCSPC) board (SPC-630, Becker and Hickl)[51].

During the experiments, the laser power was constantly maintained at ~30 ± 0.5 μW to prevent the samples from bleaching. Rhodamine B and 50 nm fluorescent microspheres were used for confocal adjustment and optical alignment, respectively. The fluorescence signal arising from the samples were collected in first-in, first-out (FIFO) with respect to the observation volume. The collected signals were first autocorrelated and then fitted by a 3-D diffusion model using Burst Analyzer 2.0 software (Becker and Hickl). The diffusion time, $\tau_D$ of the probe molecule was extracted from the autocorrelation function. Whereas, the confocal radius, $r$ was determined by measuring the diffusion coefficient ($D$) of a reference sample using Stokes-Einstein equation[52,53]. Finally, the diffusion coefficient of the probe was obtained by feeding both the $\tau_D$ and $r$ in the following equation-

$$D = \frac{r^2}{4\tau_D} \qquad (14)$$

For FCS experiments, it is mandatory to dilute the probe solution to nanomolar concentrations to maintain the presence of a single molecule in the observation volume. The higher concentration of probe adversely affects the detection of the signal. In our study, a stock solution of the fluorescent probe was prepared in water and added to the polymer samples appropriately to keep the effective concentration of the probe to 1 nM. The mixture was then sonicated for a couple of minutes to make the solution homogeneous and prevent aggregation of the probe. Subsequently 50 μL droplet of this solution was placed on a coverslip and fluorescence fluctuation was recorded 5 times for every new sample. The autocorrelation curves were ideally fitted by a one-component 3-D diffusion model, which rules out the existence of multiple sized species in the solution.

Irrespective of the Gaussian nature of the autocorrelation propagator, as long as the time interval of analysis, $\tau \ll \tau_{1/2}$ which is the estimated time taken for a probe molecule to reach to confocal beam waist or *r,* we can invert the following autocorrelation function ($G(\tau)$) to estimate 3-dimensional MSD[50]-

$$G(\tau) = \frac{1}{N}\left(1 + \frac{1}{r^2}\frac{2}{3}\,MSD\right)^{-1}\left(1 + \frac{1}{r^2 s^2}\frac{2}{3}\,MSD\right)^{-1/2} \qquad (15)$$

Where, $N$ represents the number of particles or molecules in the confocal observation volume and $s$ represents the structure factor of the observation volume or in other words, it is the ratio of confocal height ($\omega_z$) to radius (*r*), i.e. $s = \omega_z/r$.

## S6. Basic calculations and algorithm for $D_x$ estimation in PEO systems

**Varying PEO MWs system (0.5 wt%):** We coarse-grained this system into three regimes: (i) dilute (till PEO MW ~ $10^5$ where the overlap transition begins as per Kavassalis and Noolandi [54], (ii) transition (till PEO MW~$2.52 \times 10^5$ corresponding to 0.5 wt% PEO critical concentration) and (iii) semidilute (larger than PEO MW~$2.52 \times 10^5$).

**Microviscosity for small molecules:** Viscosity experienced by small molecules in polymer solution has been evaluated as per Brady's theory[55] which states that microviscosity is inversely proportional to the long-time self diffusivity in a passive microrheology regime. Following this, at lower volume fractions ($\phi \leq 0.5$), the microviscosity correlation can be evaluated as the following: $\eta_{microviscosity} \approx \eta_0(1 + 2.0\, \phi\, (1 - 0.5\phi)/(1 - \phi)^3) \sim \eta_0(1 + 2.5\, \phi + 6.2\phi^2)$. Here, $\eta_0$ is the solvent viscosity. The latter expression: $\eta_0(1 + 2.5\, \phi + 6.2\phi^2)$ is also known as Batchelor expression[56] which considers Brownian motion of hard spheres in suspension at low shear and zero-frequency conditions. Both the expressions yielded identical results in the context of our experimental systems which include small molecules and lower volume fractions of PEO.

### S6.1. Algorithm for $D_x$ estimation in varying PEO MW system

| Steps | Action |
|---|---|
| 1 | Enter the constants of the small molecule, other molecular and system properties:<br>• T- system temperature in K<br>• R-universal gas constant<br>• $N_A$-Avogadro's number<br>• $K_B$- Boltzmann's constant<br>• $k_a$- association or sticking constant of the small molecule (note: in the manuscript, it is denoted as "$K_a$")<br>• $k_D$- dissociation or escaping constant of the small molecule (note: in SI it is denoted as "$K_D$")<br>• $C^0$- 1 M<br>• $\eta_0$- viscosity of the water at temperature T<br>• $x_0$- hydrophobic decay length<br>• $a_{molecule}$- radius of the small molecule<br>• $a_{MW}$- MW of the small molecule<br>• m- mass of the small molecule<br>• $Dx_DI$-diffusion coefficient of the small-molecule in pure water at T ($D_x_DI = \frac{K_B \mathrm{T}}{6\pi\eta_0 a_{molecule}}$)<br>• $k_h$- hydrophobic adhesion force or pre-factor |
| 2 | Define polymer concentrations, molecular weights and other properties:<br>• $M_0$- molecular weight of 1 repeating monomer unit of PEO polymer<br>• $\nu$-Flory exponent in good solvent (~0.588) |
| | • $G_{N0}$-plateau modulus of PEO<br>• $b$-Kuhn's length of PEO chain<br>• $\rho_{pol}$-density of PEO polymer<br>• $N_e$- number of monomers per entanglement from $\frac{4}{5}\rho_{pol}\frac{RT}{M_0 G_{N0}}$ as per Doi-Edwards [57,58]<br>• $MW$- molecular weight of PEO<br>• $C$- conc. of PEO in wt%<br>• $C_pol$- conc. of PEO in g/ ml<br>• $N$- degree of polymerization of PEO<br>• $R_g$- radius of gyration of PEO ($R_g = 0.02\ MW^{0.58}$ )<br>• $C_{crit}$- critical concentration of PEO<br>• $a_1$-empirical exponent for evaluation of PEO solution macroviscosity [59]<br>• $a_2$- empirical exponent for evaluation of PEO solution macroviscosity [59] |
| 3 | Estimation of friction factor of PEO using Vogel-Fulcher equation: [18]<br>$$Wl^4 = Wl^4\|_\infty\ Exp(-\frac{B}{T-T_0})$$<br>• $N_{100}$- degree of polymerization of PEO with MW of 100K<br>• $l$-Segmental length of PEO chain<br>• $Wl^4$-Rouse rate at temperature T (in $nm^4/ns$)<br>• $Wl^4\|_\infty$-prefactor of eqn (16)<br>• $B$-1090 K for bulk PEO<br>• $T_0$- 155 K for bulk PEO<br>• $\zeta$- friction factor per $N$ (i.e. monomeric friction factor)<br>• $\zeta_e$- friction factor per $N_e$ (i.e. entanglement friction factor)<br>• $Wl^4\|_e(T)$- Characteristic Rouse rate at temperature T; at 348K~ 0.1 $nm^4/ns$ [60]<br>• $Wl^4\|_{e,\infty}$~ 28.36 $nm^4/ns$<br>• Calculate $\zeta_e$ as $\frac{3\ K_B T}{Wl^4\|_e(T)}\ l^2$<br>• Estimate $\zeta$ ~ $\zeta_e N^3/N_e^{3.4}$ to fit $D_x$ in semidilute regime |
| 4 | Estimation of $D_x$ in regions based on critical concentration for each MW: [61,62]<br>Estimation common to all regions:<br>• $R_t$- thermodynamic radius based on pervaded volume of the polymer ($R_t \approx 0.65R_g$)<br>• $A_2$- 2nd virial coefficient of the polymer $\frac{16}{3}\pi R_t^3\ \frac{N_A}{MW^2}$<br>• $conc_ratio$ –ratio of PEO concentration to its critical concentration ($conc_ratio = \frac{C_pol}{C_{crit}}$)<br>• $\xi_{cor}$- correlation length or mesh size of polymer chain ($\xi_{cor} = R_g\ conc_ratio^{\frac{\nu}{1-3\nu}}$)<br>• $\eta_{macroviscosity}$ – macroviscosity of PEO solution ($\eta_{macroviscosity} = \eta_0\ e^{a_2(\frac{R_g}{\xi_{cor}})^{a_1}}$) |
| | • $b_{statseg}$- statistical segment length ($b_{statseg} = \sqrt{\frac{6}{N}} R_g$ )<br><br>Dilute region ($PEO\ MW \leq\ 10^5$) :<br>• $\phi_{dil}$- polymer volume fraction in dilute region ($\phi_{dil} = \frac{C_{pol}\, N_A}{MW}\ \frac{4}{3}\pi {R_t}^3$) [62]<br>• $\eta_{theoretical}$-microviscosity followed by hard spheres ($\eta_{theoretical} = \eta_0(1 + 2.5\ \phi_{dil} + 6.2{\phi_{dil}}^2)$ ) [63]<br>• $\gamma_{theoretical}$- friction factor of small molecule ($\gamma_{theoretical} = 6\ \pi\ \eta_{theoretical}\ a_m$)<br>• $D_{x_hardsphere}$-microviscosity-driven Brownian diffusion coefficient of small molecule ($D_{x_hardsphere} = \frac{K_B \mathrm{T}}{\gamma_{theoretical}}$). The “hardsphere” term is simply indicative of no interaction in diffusion, instead of any hard sphere type repulsion.<br><br>Transition region ($10^5 \leq PEO\ MW \leq\ 2.5 \times 10^5\ i.e.\ C_pol = C_{crit}$) :<br>• $\phi_{transition}$ – polymer volume fraction at the transition between dilute and semidilute regimes ($\phi_{transition} =\ 900\ MW^{-0.77}$, which was obtained by using PEO specific constants)<br>• $\eta_{theoretical}$, $\gamma_{theoretical}$ and $D_{x_hardsphere}$ calculations are similar as described in case of dilute regime<br><br>Semi dilute region ($PEO\ MW >\ 2.5 \times 10^5$) :<br>• $\phi_{semidilute}$- polymer volume fraction in early semidilute regime, where $\phi_{semidilute} = (\frac{1}{N}\ (\frac{\eta_{macro}}{\eta_0} - 1))^{3\nu - 1}$ , and $1/(3\nu - 1)$ is the De Gennes scaling exponent [64]<br>• $\eta_{theoretical}$, $\gamma_{theoretical}$ and $D_{x_hardsphere}$ calculations are similar to dilute regime<br>• $G_N^0(\phi)$- volume fraction dependent plateau modulus in semidilute regime ($G_N^0(\phi) = G_N^0(1)\ \phi^{3\nu/(3\nu-1)}$, Where $G_N^0(1)$ is the plateau modulus of the melt as per reference [62])<br>• $atube_{theoretical}$- tube diameter ( $a_t$) evaluated using Doi-Edwards formula: $a_t^2 = \frac{4}{5}\ \frac{\rho_{pol}\, R\, T}{M_0 G_{N0}} \phi_{semidilute}\ {b_{statseg}}^2$ [58]<br>• $D_x$- diffusion coefficient of the interacting small molecule at small time interval *t* ($O(\tau_c)$):<br>$$D_x \approx\ D_{x_hardsphere}\ + \frac{1}{2}\left(\frac{x(0)}{\tau_C}\right)^2 t$$<br>Note that the characteristic interaction time $\tau_C = \frac{\zeta\ x_0}{k_h}$, where $\zeta$ is the segmental friction |
| 5 | Data plotting of selected parameters such as $D_x$ vs PEO MW |

**S6.2. Verification of volume fraction ($\phi$) correlations used:** [62,65]

Accurate estimation of polymer volume fraction is of utmost importance *w.r.t.* theoretical calculation of critical parameters of the system. We have adopted different calculation procedure of polymer volume fraction depending on the concentration regimes as follows:

**i. Dilute regime ($\phi_{dil}$):** $\phi_{dil} = C_{monomer}V_{monomer} = \left(\frac{CN_A}{M}\right)(\frac{4}{3}\pi R_t^3)$ , where $C_{monomer}$ is the monomer concentration, $V_{monomer}$ is monomer volume, $C$ is the polymer concentration, $M$ is the molecular weight, $N_A$ is the Avogadro number and $R_t$ is the thermodynamic radius of the polymer, which is roughly $R_t$~ $0.65\ R_g$ as described in S6.1. Using computational modeling results, we verified $\phi_{dil}$~ $N^{3\nu-1}$ in this regime. *N* is the degree of polymerization and $\nu$ is the Flory exponent in a good solvent.

**ii. Transition regime ($\phi_{trans} \gg \phi^*$):** The transition to semidilute regime occurs at $\phi^* = \frac{N\,b^3}{R^3} = N^{1-3\nu}$. However, the transition between dilute and semidilute regime is not abrupt as defined by $\phi^*$, but rather smooth. Using scaling theory, $\phi_{trans}$ could be derived. The end-to-end distance of polymer chain, accordingly, could be written as: $R = bN^{1/2}\phi_{trans}{}^{-0.115}$. For PEO chains, $R$~$\sqrt{6}\,R_g$~ $0.02\ MW^{\nu}$, $b$~7.1 Å, monomer molecular weight, $M_0$~ 44. Taking all the parameter values together, we could write $\phi_{trans}$~$900\ MW^{-0.76}$. Using computational modeling results, we verified $\phi_{dil}$~ $N^{1-3\nu}$ in this regime.

**iii. Semidilute regime ($\phi_{semi}$):** Since solution viscosity is well correlated to PEG or PEO molecular weights, we used specific viscosity ($\eta_{sp}$) dependent volume fraction expression as per Rubinstein [66]. The expression of volume fraction in the semi-dilute regime ($\phi_{semi}$) could be derived from the following relation: $\eta_{sp} = N\ \phi_{semi}{}^{\frac{1}{3\nu-1}}$ , where $\frac{1}{3\nu-1}$ is the De Gennes scaling exponent. Using computational modeling results and power law fittings, we verified $\phi_{semi}$~ $N^{2(3\nu-1)}$ which follows the dynamics of entangled chains.

**S6.3. Entangled regime definition according to Kavassalis-Noolandi transition parameter and other entanglement criteria:**

According to Kavassalis and Noolandi transition parameter[54], the larger the *N*, the earlier the transition overlap *w.r.t.* $\phi$. In case of good solvent, with $\phi$~0.1, the overlap initiation takes place roughly at PEO MW ~ $10^5$ ($N$~$2.2 \times 10^3$) in our computational demonstration. Using the number of monomer repeats between two successive entanglements in PEO chains i.e. $N_e$~32, we estimated the Kavassalis-Noolandi coordination parameter: $\tilde{N}$+1 ~ 4 (here we assumed all other parameters equivalent to polyethylene chains). Using the above estimation of $\tilde{N}$+1, the critical volume fraction for entanglement transition, $\phi_c$~$3.6 \times 10^{-3}$ for PEO MW of 200K. Interestingly, with $\phi$~$3.3 \times 10^{-3}$, 0.5 wt% PEO 200K network becomes a borderline case of entanglement initiation. In the case of PEO MW of 500K, the critical volume fraction for entanglement transition, $\phi_c$~$1.7 \times 10^{-3}$ which is satisfied by 0.5 wt% PEO 500K network in our computational demonstration. Therefore, the reported MWs (Figure 2c and Figure 5b) in the semidilute regime are all entangled chains.

This could be further corroborated by higher than an order of magnitude ratio of polymer overlap parameter ($O$) to the Kavassalis-Noolandi number defined as $O_{KN}$ ($O_{KN}$~ $11, O/O_{KN} \gg 1$). According to this theory, if overlap parameter $O < O_{KN}$ then linear chains, topologically, should not restrict the motion of neighboring chains, which implies the absence of entanglement. Chain entanglement is expected for high MW linear PEO chains with $O_{KN} \geq 10$ and higher than an order of magnitude ratio of $O$ to $O_{KN}$ i.e. $O/O_{KN} \gg 1$[67]. These parameter values indicate significant chain overlap and activation of topological effects as depicted by tube structures.

Additionally, for 0.5 wt% solution of PEO MW 500K, the critical concentration in g/ml- $C^{*} \sim N^{-4/5} \sim 5 \times 10^{-4}$, which brings the ratio of $C/C^{*} \sim (5 \times 10^{-3})/\ (5 \times 10^{-4}) \sim 10$ and therefore, $C$ would be in entangled regime [64]. Similarly, the ratio of solution viscosity at $C$ in case of 500K PEO solution to solvent (water) viscosity is also ~ 10 as per one of the entanglement criteria [64].